\documentclass[twocolumn,longauth]{aa}

\usepackage{caption}
\usepackage{graphicx}
\usepackage{txfonts}
\usepackage{lipsum}
\usepackage{subcaption}         
\usepackage{lscape}             
\usepackage{placeins}           
\usepackage{booktabs}

\usepackage[]{hyperref}

\begin{document}
	
	\title{\texttt{photoD} with Rubin's Data Preview 1: first stellar photometric distances and deficit of faint blue stars}
	
	\subtitle{Stellar distances with Rubin's DP1}
	
		
\author{
	L.~Palaversa\inst{1}\fnmsep\thanks{\email{lp@irb.hr}}
	\and E.~Donev\inst{2}
	\and Ž.~Ivezi\'{c}\inst{3,4}
	\and K.~Mrakov\v{c}i\'{c}\inst{5}
	\and N.~Caplar\inst{4}
	\and M.~Juri\'{c}\inst{6}
	\and T.~Jurki\'{c}\inst{5}
	\and S.~Campos\inst{7}
	\and M.~DeLucchi\inst{7}
	\and D.~Jones\inst{6}
	\and K.~Malanchev\inst{7}
	\and A.~I.~Malz\inst{7}
	\and S.~McGuire\inst{7}
	\and B.~Abel\inst{8}
	\and L.~Girardi\inst{9}
	\and G.~Pastorelli\inst{9,10}
	\and M.~Trabucchi\inst{10}
	\and S.~Zaggia\inst{9}
	\and E.~Acosta\inst{3}
	\and C.~L.~Adair\inst{11}
	\and J.~Andrew\inst{3}
	\and É.~Aubourg\inst{12}
	\and A.~E.~Bauer\inst{13}
	\and W.~Beebe\inst{6}
	\and E.~C.~Bellm\inst{4}
	\and R.~D.~Blum\inst{14}
	\and M.~T.~Booth\inst{3}
	\and A.~Boucaud\inst{15}
	\and D.~Branton\inst{6}
	\and D.~L.~Burke\inst{16}
	\and D.~Calabrese\inst{3}
	\and J.~L.~Carlin\inst{3}
	\and H-F.~Chiang\inst{11}
	\and Y.~Choi\inst{17}
	\and A.~J.~Connolly\inst{6}
	\and S.~Dagoret-Campagne\inst{18}
	\and P.~N.~Daly\inst{3}
	\and F.~Daruich\inst{19}
	\and G.~Daubard\inst{20}
	\and E.~Dennihy\inst{3}
	\and H.~Drass\inst{19}
	\and O.~Eiger\inst{11,16}
	\and A.~M.~Eisner\inst{21}
	\and L.~P.~Guy\inst{19}
	\and J.~Hoblitt\inst{3}
	\and P.~Ingraham\inst{22}
	\and F.~Jammes\inst{23}
	\and B.~T.~Jannuzi\inst{24}
	\and M.~J.~Jee\inst{25,26}
	\and T.~Jenness\inst{3}
	\and R.~L.~Jones\inst{4}
	\and C.~Juramy-Gilles\inst{20}
	\and S.~M.~Kahn\inst{27}
	\and Y.~Kang\inst{16,19}
	\and A.~Kannawadi\inst{28,29}
	\and L.~S.~Kelvin\inst{29}
	\and I.~V.~Kotov\inst{30}
	\and G.~Kov\'acs\inst{6}
	\and N.~R.~Kurita\inst{11}
	\and T.~Lange\inst{11}
	\and D.~Laporte\inst{20}
	\and J.~C.~Lazarte\inst{11}
	\and S.~Liang\inst{11}
	\and M.~Lopez\inst{11}
	\and N.~B.~Lust\inst{29}
	\and M.~Lutfi\inst{3}
	\and O.~Lynn\inst{7}
	\and G.~Mainetti\inst{31}
	\and F.~Menanteau\inst{32}
	\and M.~Miller\inst{17}
	\and M.~Moniez\inst{18}
	\and N.~Sedaghat\inst{4}
	\and E.~Nourbakhsh\inst{29}
	\and H.~Y.~Park\inst{28}
	\and J.~R.~Peterson\inst{33}
	\and R.~Plante\inst{34}
	\and A.~A.~Plazas~Malag\'on\inst{11,16}
	\and M.~N.~Porter\inst{35}
	\and K.~A.~Reil\inst{11}
	\and V.~J.~Riot\inst{36}
	\and A.~Roodman\inst{16}
	\and E.~S.~Rykoff\inst{16}
	\and R.~H.~Schindler\inst{16}
	\and J.~Sebag\inst{19}
	\and R.~A.~Shaw\inst{37}
	\and A.~Shugart\inst{19}
	\and K.~B.~Siruno\inst{19}
	\and J.~A.~Smith\inst{38}
	\and J.~D.~Swinbank\inst{39,29}
	\and J.~G.~Thayer\inst{11}
	\and S.~Thomas\inst{3}
	\and R.~Tighe\inst{19}
	\and D.~L.~Tucker\inst{40}
	\and M.~Turri\inst{11}
	\and E.~K.~Urbach\inst{41}
	\and B.~Van~Klaveren\inst{11}
	\and W.~van~Reeven\inst{19}
	\and C.~Z.~Waters\inst{29}
	\and B.~Willman\inst{42}
}
\institute{
	Ru{\dj}er Bo{\v{s}}kovi{\'c} Institute, Bijeni{\v{c}}ka cesta 54, 10000 Zagreb, Croatia
	\and XV. Gymnasium (MIOC), Jordanovac 8, 10000, Zagreb, Croatia
	\and Vera C.\ Rubin Observatory Project Office, 950 N.\ Cherry Ave., Tucson, AZ  85719, USA
	\and University of Washington, Dept.\ of Astronomy, Box 351580, Seattle, WA 98195, USA
	\and Faculty of Physics, University of Rijeka, Radmile Matej\v{c}i\'{c} 2, Rijeka, Croatia
	\and Institute for Data-intensive Research in Astrophysics and Cosmology, University of Washington, 3910 15th Avenue NE, Seattle, WA 98195, USA
	\and McWilliams Center for Cosmology \& Astrophysics, Department of Physics, Carnegie Mellon University, Pittsburgh, PA 15213, USA
	\and Olympic College, 1600 Chester Ave., Bremerton, WA 98337-1699, USA
	\and INAF - Osservatorio Astronomico di Padova, Vicolo dell`{O}sservatorio 5, I-35122 Padova
	\and Department of Physics and Astronomy G. Galilei, University of Padova, Vicolo dell’Osservatorio 3, I-35122, Padova, Italy
	\and SLAC National Accelerator Laboratory, 2575 Sand Hill Rd., Menlo Park, CA 94025, USA
	\and Universit\'{e} Paris Cit\'{e}, CNRS/IN2P3, CEA, APC, 4 rue Elsa Morante, F-75013 Paris, France
	\and Yerkes Observatory, 373 W. Geneva St., Williams Bay, WI 53191, USA
	\and Vera C.\ Rubin Observatory/NSF NOIRLab, 950 N.\ Cherry Ave., Tucson, AZ  85719, USA
	\and Universit\'{e} Paris Cit\'{e}, CNRS/IN2P3, APC, 4 rue Elsa Morante, F-75013 Paris, France
	\and Kavli Institute for Particle Astrophysics and Cosmology, SLAC National Accelerator Laboratory, 2575 Sand Hill Rd., Menlo Park, CA 94025, USA
	\and NSF NOIRLab, 950 N.\ Cherry Ave., Tucson, AZ 85719, USA
	\and Universit\'{e} Paris-Saclay, CNRS/IN2P3, IJCLab, 15 Rue Georges Clemenceau, F-91405 Orsay, France
	\and Vera C.\ Rubin Observatory, Avenida Juan Cisternas \#1500, La Serena, Chile
	\and Sorbonne Universit\'{e}, Universit\'{e} Paris Cit\'{e}, CNRS/IN2P3, LPNHE, 4 place Jussieu, F-75005 Paris, France
	\and Santa Cruz Institute for Particle Physics and Physics Department, University of California--Santa Cruz, 1156 High St., Santa Cruz, CA 95064, USA
	\and Steward Observatory, The University of Arizona, 933 N.\ Cherry Ave., Tucson, AZ 85721, USA
	\and Universit\'{e} Clermont Auvergne, CNRS/IN2P3, LPCA, 4 Avenue Blaise Pascal, F-63000 Clermont-Ferrand, France
	\and University of Arizona, Department of Astronomy and Steward Observatory, 933 N.\ Cherry Ave, Tucson, AZ 85721, USA
	\and Department of Astronomy, Yonsei University, 50 Yonsei-ro, Seoul 03722, Republic of Korea
	\and Physics Department, University of California, One Shields Avenue, Davis, CA 95616, USA
	\and Physics Department,  University of California, 366 Physics North, MC 7300 Berkeley, CA 94720, USA
	\and Department of Physics, Duke University, Durham, NC 27708, USA
	\and Department of Astrophysical Sciences, Princeton University, Princeton, NJ 08544, USA
	\and Brookhaven National Laboratory, Upton, NY 11973, USA
	\and CNRS/IN2P3, CC-IN2P3, 21 avenue Pierre de Coubertin, F-69627 Villeurbanne, France
	\and NCSA, University of Illinois at Urbana-Champaign, 1205 W.\ Clark St., Urbana, IL 61801, USA
	\and Department of Physics and Astronomy, Purdue University, 525 Northwestern Ave., West Lafayette, IN  47907, USA
	\and National Institute of Standards and Technology, Material Measurement Laboratory, Office of Data and Informatics, Gaithersburg, MD, USA
	\and Florida Institute of Technology, 150 W. University Blvd., Melbourne, FL 32901, USA
	\and Lawrence Livermore National Laboratory, 7000 East Avenue, Livermore, CA 94550, USA
	\and Space Telescope Science Institute, 3700 San Martin Drive, Baltimore, MD 21218, USA
	\and Austin Peay State University, Clarksville, TN 37044, USA
	\and ASTRON, Oude Hoogeveensedijk 4, 7991 PD, Dwingeloo, The Netherlands
	\and Fermi National Accelerator Laboratory, P. O. Box 500, Batavia, IL 60510, USA
	\and Department of Physics, Harvard University, 17 Oxford St., Cambridge MA 02138, USA
	\and LSST Discovery Alliance, 933 N. Cherry Ave., Tucson, AZ 85719, USA
}

	\date{\today}
	\abstract
	{}
        {We investigate the utility of Rubin's Data Preview 1 for estimating stellar number density profile in the Milky Way halo.}
	{Stellar broad-band near-UV to near-IR $ugrizy$ photometry released in Rubin’s Data Preview 1 
	is used to estimate distance and metallicity for blue main sequence stars 
	brighter than $r=24$ in three $\sim$1.1 sq.~deg. fields at southern Galactic latitudes.}
	{Compared to TRILEGAL simulations of the Galaxy's stellar content by \cite{dal_tio_simulating_2022}, we find a significant deficit of
	blue main sequence turn-off stars with $22 < r < 24$. We interpret this discrepancy as a signature of a much steeper halo number
	density profile at galactocentric distances $10-50$ kpc than the cannonical $\sim1/r^3$ profile assumed in TRILEGAL simulations.}
	{This interpretation is consistent with earlier suggestions based on observations of more luminous, but much
	less numerous, evolved stellar populations, and a few pencil beam surveys of blue main sequence stars in the
	northern sky. These results bode well for the future Galactic halo exploration with Rubin's Legacy Survey of Space and Time.}	
	
	\maketitle
	   \nolinenumbers

\section{Introduction} \label{sec:intro}

Thanks to the advent of sensitive wide-area digital sky surveys, the last two decades have seen tremendous
progress in mapping of the Galaxy's principal components: the bulge, thin and thick disks and the halo
\cite[e.g.,][]{1986ARA&A..24..577B, 2012ARA&A..50..251I, 2021ApJ...911..109S, 2024ApJ...966..159F, 2024ApJ...975...81Y, 2024MNRAS.531.4762M, 2025PASJ...77..178F}. 
Studies of the Galactic stellar halo provide especially powerful clues for the  Galaxy's formation and evolution 
history due to its long dynamical time scales. This region also provides unique constraints for the Galaxy's total mass and extent,
which are important parameters for near-field cosmology. 

The distant stellar halo profile was first measured beyond a galactocentric distance of $r_{gc}\sim$10 kpc by \cite{1996AJ....112.1046W}.
They analyzed the distribution of RR Lyrae stars and estimated that the stellar number density profile (hereafter the radial profile)
follows a $1/r^n_{gc}$ power law with $n\sim$3 out to $\sim$80 kpc. More recent studies of RR Lyrae stars and other luminous tracers, such as blue horizontal
branch stars and M giants, found evidence that the radial profile steepens beyond about 30 kpc
\cite[e.g.,][]{2014ApJ...788..105F, 2022AJ....164..249H, 2024ApJ...975...81Y, 2025PASJ...77..178F}. For example, Figure 7 in
\cite{2024MNRAS.531.4762M} visually summarizes the radial profile vs. $r_{gc}$ results from many studies and shows
that at about $r_{gc}\sim$ $20-30$ kpc, the profile power-law index $n$ changes from about 2-3 to about 4-5. Similar conclusions were
drawn using main sequence stars (typically blue turn-off stars) observed in a few small-area pencil beam surveys
\citep[e.g.,][]{2011ApJ...731....4S}, and from a global fitting of colour-magnitude diagrams down to g=23 from the Dark Energy Survey, covering ~2300 deg$^2$ \citep{2020MNRAS.497.1547P}. These findings are important for understanding the accretion history of the Milky Way.
For example, based on a comparison to numerical simulations, \cite{Deason_2014} found that stellar halos with
shallower slopes at large distances tend to have more recent accretion activity and concluded that steeper observed slopes
suggest that the Milky Way has had a relatively quiet accretion history over the past several gigayears.  

However, \cite{2024MNRAS.531.4762M} emphasized that results for the radial profile seem to depend on the regions of the sky
surveyed, implying that for robust conclusions large sky areas must be examined. Furthermore, the spatial resolution for
mapping the halo can be greatly enhanced by studying main sequence stars which are much more numerous than the more luminous
stellar populations. In order to explore the halo to distances of $\sim$100 kpc
(distance modulus of 20 mag) with turn-off stars (absolute magnitude $M_r \sim 5$), a sky survey with reliable star-galaxy separation
performance to $r\sim25$ is needed. With ground-based seeing of the order 1 arcsec, the 5$\sigma$ limiting depth of $r \sim 27$ is
required to ensure such performance due to galaxies vastly outnumbering stars at such faint apparent magnitudes \citep{2020AJ....159...65S}. 
Therefore, a large-area sky survey with the $5\sigma$ limiting depth of $r\sim27$ is required to explore the distant halo with numerous
main-sequence stars. 

The upcoming Legacy Survey of Space and Time (LSST) about to be started at the NSF-DOE Vera C. Rubin Observatory \citep{2019ApJ...873..111I}
meets these requirements for sky coverage and depth. Over its main 18,000 deg$^2$. footprint, LSST is expected to reach a depth
of $r\mathord{\sim}27$ \citep{2022ApJS..258....1B}. The Rubin Observatory has recently released Data Preview 1 \citep[][hereafter DP1]{RTN-095}
-- its first public dataset based on commissioning observations with an engineering camera \citep{10.71929/rubin/2561361}.
While much smaller and shallower, except in the case of the Extended Chandra Deep Field South (ECDFS), than the ultimate LSST dataset, DP1 presents the first opportunity to assess the potential
of LSST, to test its data quality and processing pipeline performance, as well as various tools that are being developed to
distribute and analyze LSST data. 

One such tool is a Bayesian software framework for rapid estimation of distance, metallicity and interstellar dust extinction along
the line of sight for stars in the Galaxy \citep{2025AJ....169..119P}. This framework, named \texttt{photoD}, will be applied to future LSST
data releases and here we provide the first test of its performance on real Rubin data. 
We briefly overview our methodology and describe the DP1 dataset in \S2. We present our analysis in \S3, and
summarize and discuss our results and possibilities for further improvements in \S4.

\section{Methodology}
\label{sec:method}

In this Section we describe the data used in our analysis, selection of the final stellar samples, and the method
used for estimating stellar distances and photometric metallicity.

\subsection{Rubin Observatory's Data Preview 1 Photometry}

\begin{figure*}[h!]
\hskip 0.1in
\includegraphics[width=0.95\textwidth,angle=0]{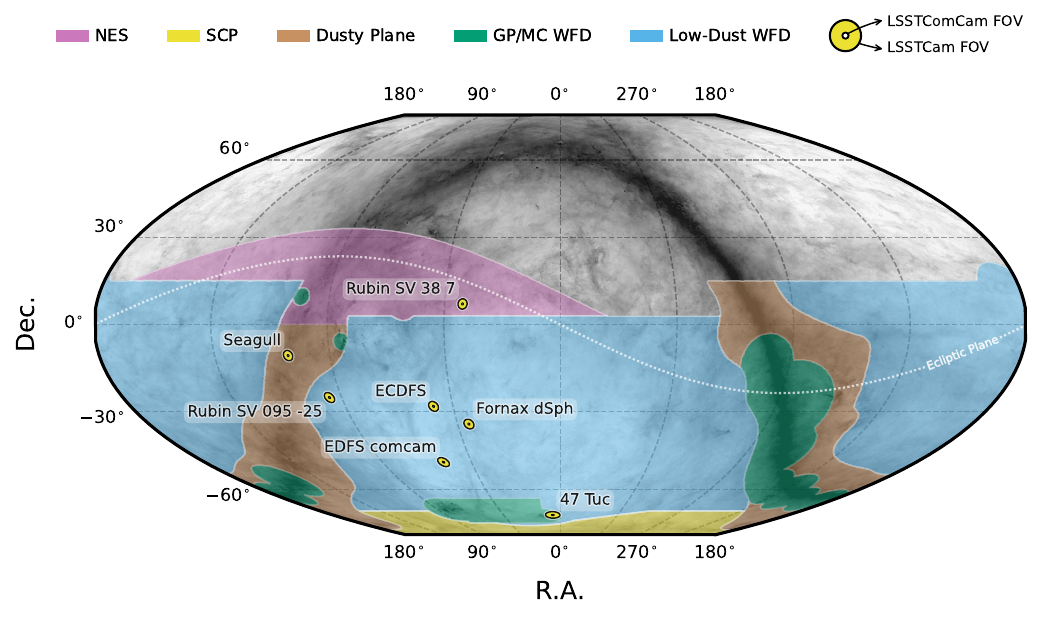}
\caption{The sky coverage of Rubin Data Preview 1 dataset. The seven LSSTComCam fields are shown as yellow dots (with names)
in the context of the LSST's planned regions: North Ecliptic Spur (purple), South Celestial Pole (yellow), the dusty plane, the Galactic
Plane and Magellanic Clouds (brown and green), and low-dust-extinction regions of the Wide-Fast-Deep (blue) program. Here we analyze
data from three fields: ECDFS, EDFS and Rubin SV 95 -25. Adapted from \citet{10.71929/rubin/2570308} (\url{https://dp1.lsst.io/}).}
\label{fig:DP1fields}
\end{figure*}


Several papers have already utilized DP1 for galactic studies \citep{2025arXiv250701343C, 2025RNAAS...9..171W, 2025RNAAS...9..161C, 2025arXiv250623955M}.
We used the Rubin Science Platform's Table Access Protocol to obtain
\texttt{b\_free\_psfFlux}\footnote{Flux derived from using the PSF
  model as a weight function and measured on band \textit{b}.}. fluxes from the \texttt{Object} catalog for each of the seven DP1 fields. Each of these fields covers an area of approximately 1.1 deg$^2$. and their locations are illustrated in Fig.\ref{fig:DP1fields}. The released data contain observations taken from October 24
to December 11, 2024, and include a total of about 2,000 30- or 38-second exposures taken through Rubin \textit{ugrizy} filters by the engineering camera (LSSTComCam)\footnote{Exposures in the \textit{u}-band are 38-second long as the sky background noise is more easily dominated by read noise.}.
The LSSTComCam contains a single "raft" of nine 4k x 4k CCDs arranged in a 3x3 square grid and placed in the center of the Simonyi Survey Telescope's field of view. The plate scale of the images is equal to that of the full LSSTCam, however, the total field of view is significantly smaller (21 times). 

Because of the commissioning requirements, the cadence of the observations was significantly different for each of the fields. Additionally the 47 Tuc, Rubin SV 38 7, Fornax dwarf spheroidal galaxy and Seagull nebula fields had less than 10 visits in either \textit{u} or \textit{i} filters. Therefore, we excluded those fields from our analyses and focused our investigations on the three best-observed fields: ECDFS, Euclid Deep Field South (EDFS) and Rubin SV 95 $-$25.
Further details about these fields, such as sky locations and the number of exposures per band, are listed in Tables \ref{tab:fields} and \ref{tab:depths}. 
The integrated depth of an unresolved source at the signal-to-noise ratio of five (5$\sigma$) is in the 26.0 $\lesssim r \lesssim 26.8$ range for the three fields studied in this paper (approximately, the ECDFS depth is comparable to a 10-year LSST survey while the other two fields correspond to about 3-4 years of LSST data). 
Selection of stellar candidates from these catalogs is discussed below. 

\begin{table*}[t]
	\centering
	\small
	\caption{Summary of basic characteristics for the three Rubin DP1 fields analyzed in this work, as well as counts of objects after successive selection steps. Equatorial and galactic coordinates are listed in degrees. The $ugrizy$ columns list the number of exposures in each band. The last four columns list the number of objects after successive selection steps. Each field covers about 2 deg$^2$.}
	\label{tab:fields}
	\begin{tabular}{lcccccccccccccc}
		\toprule
		\textbf{Field} &
		\multicolumn{4}{c}{\textbf{Coordinates}} &
		\multicolumn{6}{c}{\textbf{The number of exposures per band}} &
		\multicolumn{4}{c}{\textbf{Selected sample sizes}} \\
		\cmidrule(lr){2-5}
		\cmidrule(lr){6-11}
		\cmidrule(l){12-15}
		& R.A. & Dec & $l$ & $b$ &
		$u$ & $g$ & $r$ & $i$ & $z$ & $y$ &
		Total & $r<24$ & Unresolved & Blue MS \\
		\midrule
		ECDFS                 & 53.13 & $-$28.10 & 224.07 & $-$54.47  &  43 & 230 & 237 & 162 & 153 & 30 & 494851 &  81622   &   6874 &   786 \\
		EDFS                   & 59.10 & $-$48.73 & 256.98 & $-$48.48 &  20 & 61  & 87  & 42  & 42  & 20 & 375663 &  83868   &   8454 & 1010 \\
		SV 95$-$25      & 95.00 & $-$25.00 & 232.53 & $-$17.64  &  33 & 82  & 84  & 23  & 60  & 10 & 348913 & 128130  & 30393 & 5767 \\
		\bottomrule
	\end{tabular}
\end{table*}

\begin{table*}
	\caption{Median $5\sigma$ coadded point-source detection limits per field and band. Adapted from \citep{RTN-095}.}
	\label{tab:depths}
	\centering
	\begin{tabular}{lcccccc}
		\toprule
		Field & $u$ & $g$ & $r$ & $i$ & $z$ & $y$ \\
		\midrule
		ECDFS & 24.55 & 26.18 & 25.96 & 25.71 & 25.07 & 23.10 \\
		EDFS  & 23.42 & 25.77 & 25.72 & 25.17 & 24.47 & 23.14 \\
		SV 95$-$25 & 24.29 & 25.46 & 24.95 & 24.86 & 24.32 & 22.68 \\
		\bottomrule
	\end{tabular}
\end{table*}

\subsection{Additional photometry from extant catalogs\label{sec:extant}}

To validate the results obtained from the Rubin DP1, we utilize photometric catalogs derived from observations made by the Dark Energy Camera (DECam) on the 4-meter Blanco Telescope. Corresponding catalogs were published as part of the Dark Energy Survey Data Release 2 \citep[DES][]{desdr2} and the DECam Local Volume Exploration Survey Data Release 2 \citep[DELVE][]{delvedr2}. These surveys provide the deepest optical observations of the three DP1 fields investigated in this paper. 

The DES Data Release 2 catalog is derived from coadded images assembled from 6 years of observations and covers about 5000 deg$^2$ in the \textit{grizY} bands. Its sky coverage includes Rubin DP1 ECDFS and EDFS fields. In the filters relevant to this work, the median coadded catalog depth in a $1.95$ arcsec aperture at S/N=5 is \textit{g}=25.4 , \textit{r}=25.1 and \textit{i}=24.5. The photometric accuracy is estimated at approximately 11 mmag and the photometric uniformity has a standard deviation less than 3 mmag. To select high-confidence photometry and separate extended and point-like sources we require for each \textit{gri} band (\textit{b}) that \texttt{flags\_{b} $<$ 4}, \texttt{imaflags\_iso\_{b} == 0} and \texttt{extended\_class\_coadd $<=$ 0}.

The DELVE Data Release 2 combines public archival DECam data with more than 150 nights of additional observations for a total coverage of more than 21,000 deg$^2$ of the high-Galactic-latitude sky that includes the Rubin SV 95 $-$25 field. The median PSF depth for point like sources in the \textit{g}, \textit{r}, \textit{i} bands at S/N=5 is estimated at 24.2, 23.8 and 23.4 mag, respectively, while the absolute photometric uncertainty is $\lessapprox$ 20 mmag. To separate extended and point-like sources we require that the \textit{gri} bands simultaneously satisfy the condition \texttt{EXTENDED\_CLASS\_{b} == 0}, where \textit{b} stands for band.

\subsection{Star-galaxy Separation with Data Preview 1 Photometry \label{sec:selection}}

Selection of unresolved sources, popularly known as star-galaxy separation\footnote{At the faint magnitudes
probed here, many galaxies and quasars are unresolved in ground-based seeing (here in the range 1.0-1.3 arcsec). In addition, even some stars
may appear resolved (for example, stars with circumstellar shells and close binary stars). Therefore, ``star-galaxy separation'' really means
separation of unresolved and resolved sources. For more details, please see \cite{2020AJ....159...65S}.}, is an important methodological step
because its limitations can result in both incomplete and contaminated samples of unresolved objects,
presumably stars. These difficulties are exacerbated at the faint magnitudes probed by Rubin because galaxies
can outnumber stars by more than an order of magnitude at high galactic latitudes. 

The Rubin DP1 Object and Source catalogs include \texttt{extendedness}, a binary quantity provided in each bandpass. It is set to unity when the point spread function (PSF) magnitude exceeds the CModel magnitude\footnote{The CModel (or "composite model") flux is computed with a best-fit source surface brightness profile, while PSF flux is computed with the point spread function profile.
For extended sources, the PSF flux is biased low. For more details, see \cite{2020AJ....159...65S}.} by 0.016 mag,
indicating statistical evidence for a resolved source. The statistical properties of this ``star-galaxy separator'', 
and its comparison to analogous quantities, such as \texttt{EXTENDED\_CLASS} used by the DES pipelines,
are discussed in detail by \cite{2020AJ....159...65S}. 

Given that \texttt{extendedness} values in each bandpass are measured nearly independently, they
can be combined to increase the accuracy of star-galaxy separation. The optimal probabilistic approach
is described in \cite{2020AJ....159...65S} and relies on a quantitative description of the source distribution
in the CModel vs. PSF magnitude space, as well as on adequate Bayesian priors. Since neither is yet available for
the DP1 dataset, we employ an ansatz to combine measurements from the $gri$ bandpasses, as follows.

\begin{figure}[h!]
\includegraphics[width=0.495\textwidth,angle=0]{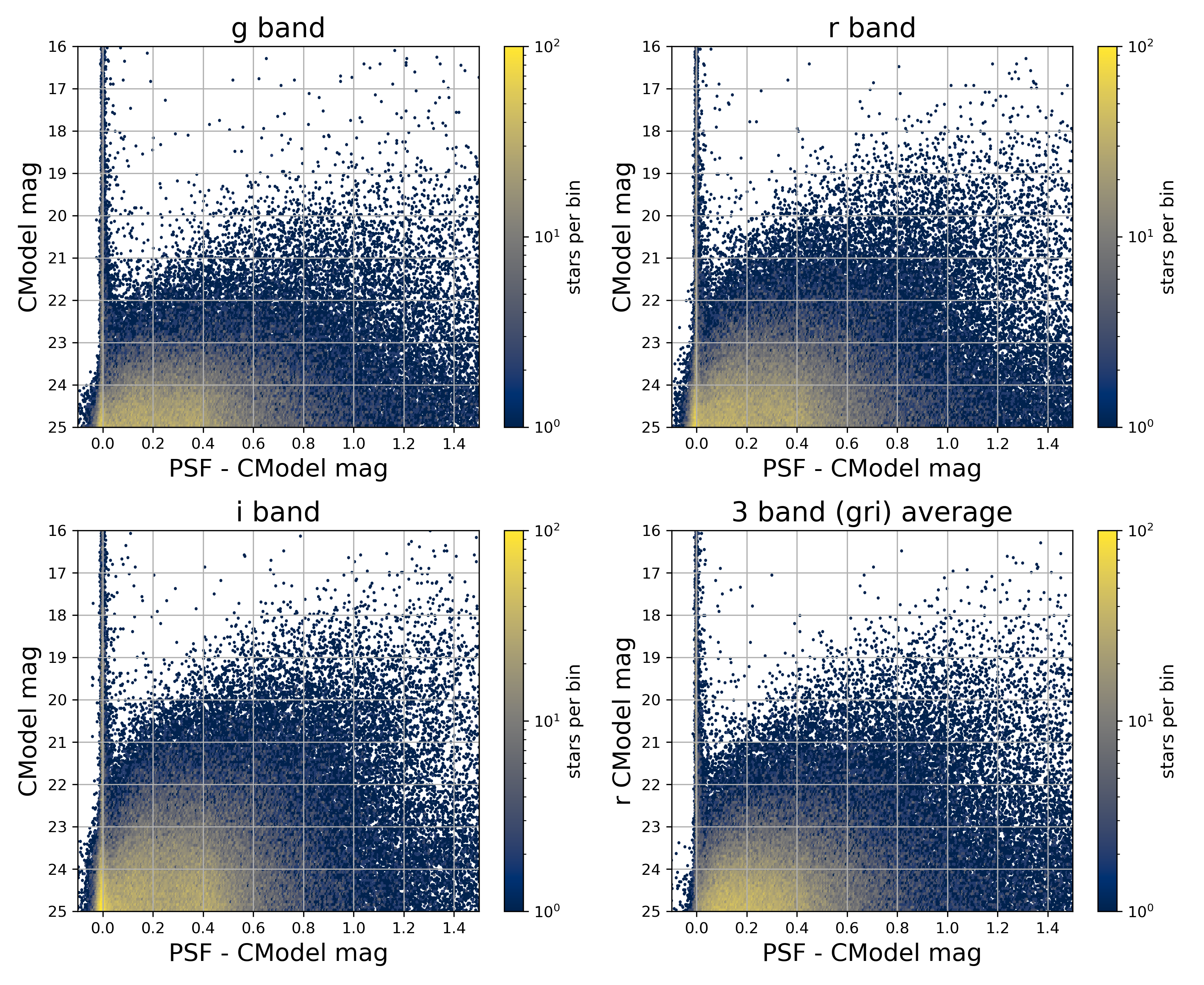}
\caption{Difference between the PSF (point spread function) magnitude and the CModel magnitude in the
  $gri$ bands, and the mean $gri$ value (bottom right) vs. magnitude for the ECDFS field (analogous to Figure 1 from
  \citealt{2020AJ....159...65S}).  The colormap encodes the number of stars per bin. The bimodal
  distribution of stars and galaxies (more precisely, unresolved and resolved sources) is evident.} 
\label{fig:SG}
\end{figure}

\begin{figure}[h!]
\includegraphics[width=0.495\textwidth,angle=0]{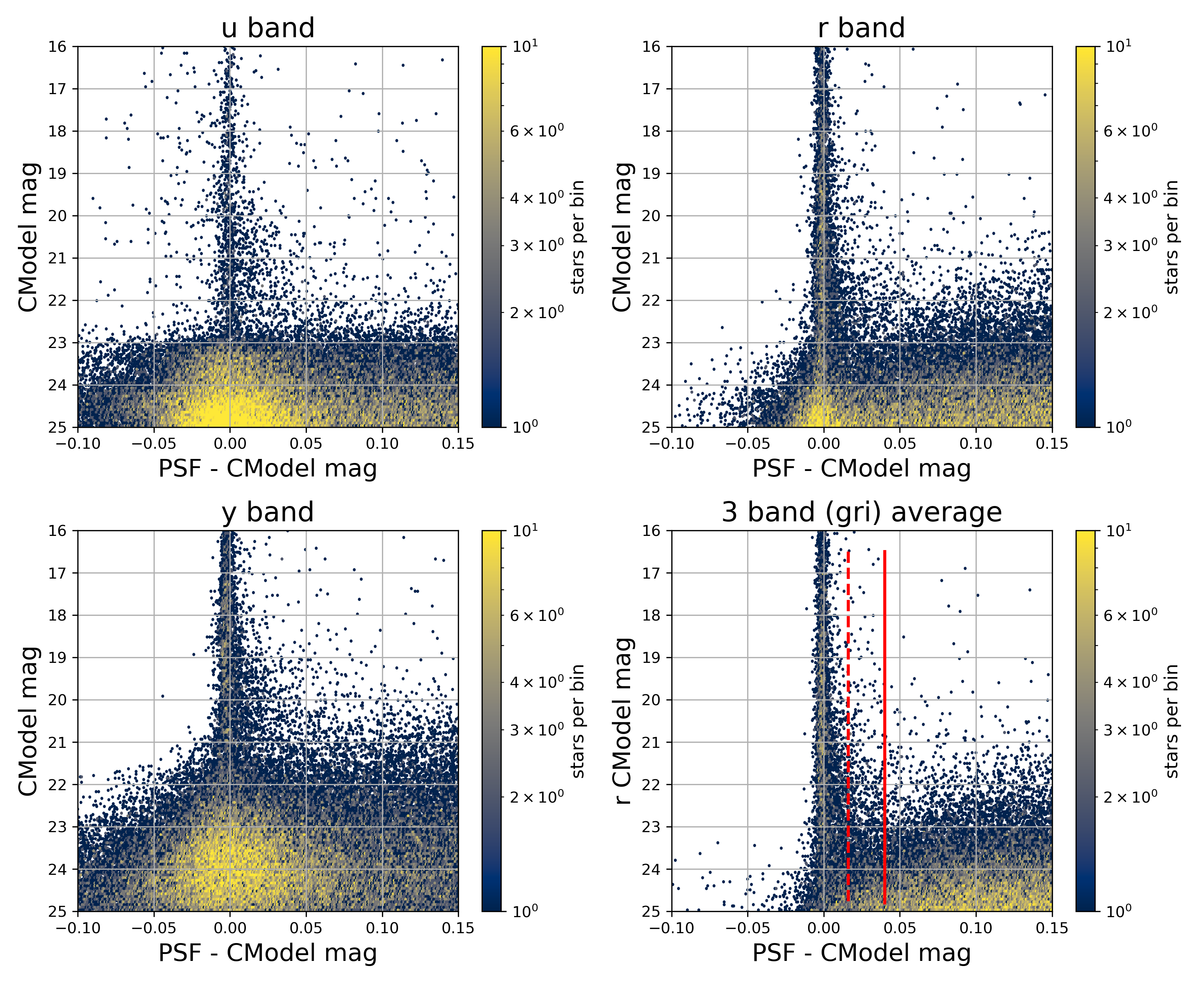}
\caption{Analogous to Figure~\ref{fig:SG}, except that the abscissa is zoomed in on, and the left panels show the $u$ and
  $y$ band diagrams, respectively.  The two vertical lines in the bottom right panel show the separation boundary between unresolved
  and resolved sources (dashed at 0.016 mag: Rubin default value for single-band classification; solid at 0.04 mag: adopted here for the
  mean $gri$ values and designed to ensure stellar completeness to faint magnitude limits).}
\label{fig:SGzoom}
\end{figure}

Figure~\ref{fig:SG} shows the variation of the difference of the CModel and PSF magnitudes as a function of
magnitude. The distribution of sources in this diagram is similar in the three shown bands, that have the
highest signal-to-noise ratio, and we take the mean of the three values. This averaging operation results
in a somewhat narrower distribution between the difference of CModel and PSF magnitudes for the vertical plume
corresponding to unresolved sources, though it is not possible to precisely quantify the improvement in
star-galaxy separation without adequate training samples. Using the mean value, we select candidates
for unresolved sources by limiting the magnitude difference to 0.04 mag (see the bottom right panel in
Figure~\ref{fig:SGzoom}). The adopted value is larger than the default value of 0.016 mag adopted by Rubin
pipelines and is motivated by the morphology of the source distribution. We selected a criterion more inclusive than
the default because we will discuss below a \textit{deficit} of faint blue stars. In other words, we chose to err on the
side of galaxy contamination in the stellar sample, rather than stellar sample incompleteness. 

\begin{figure}[h!]
	\includegraphics[width=0.495\textwidth,angle=0]{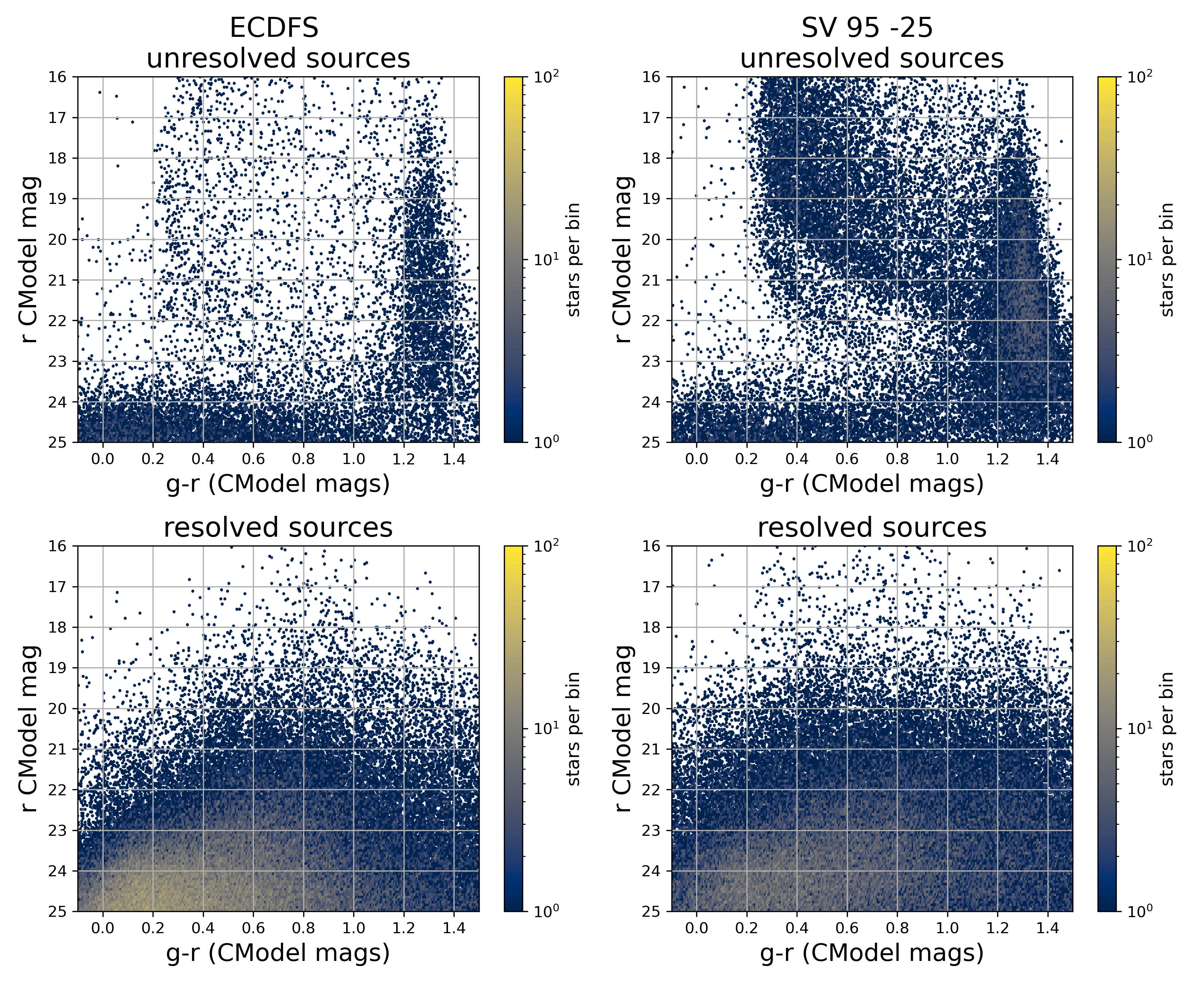}
	\caption{A comparison of color-magnitude diagrams of unresolved (top) and resolved (bottom) sources selected from Rubin DP1
          fields ECDFS (left) and SV 95 $-$25 (right). Note the low fractions of unresolved sources (approximately 8\% in the ECDFS
          field and 24\% in the SV 95 $-$25 field, for $r<25$).}
	\label{fig:compareCMDs}
\end{figure}

Using this method, we construct separate color-magnitude diagrams for unresolved and resolved sources
shown in Figure~\ref{fig:compareCMDs}.  The morphology of the source distribution in those diagrams
clearly indicates that the much less numerous stellar sample becomes contaminated by galaxies for
$r > 24$. This contamination could be alleviated somewhat statistically with appropriate use of priors
\citep{2020AJ....159...65S}.

Priors are usually computed using simulations of the Milky Way stellar content, such as TRILEGAL
\citep{dal_tio_simulating_2022}. TRILEGAL is a state-of-the-art simulation of anticipated LSST stellar content
and is publicly accessible via NOIRLab's Astro Data Lab (\url{https://datalab.noirlab.edu}).
We report here a detection of a faint cutoff in the distribution of blue stars ($g-r < 0.6$) at $r\sim22$ that
is in strong conflict with predictions of the TRILEGAL model. For this reason, using TRILEGAL-based priors
in the star-galaxy separation step seems premature. We proceed to discuss this empirical finding of a 
deficit in the counts of faint blue stars in more detail. 

\subsection{Deficit of Faint Blue Stars}

\begin{figure*}[h!]
	\includegraphics[width=0.99\textwidth,angle=0]{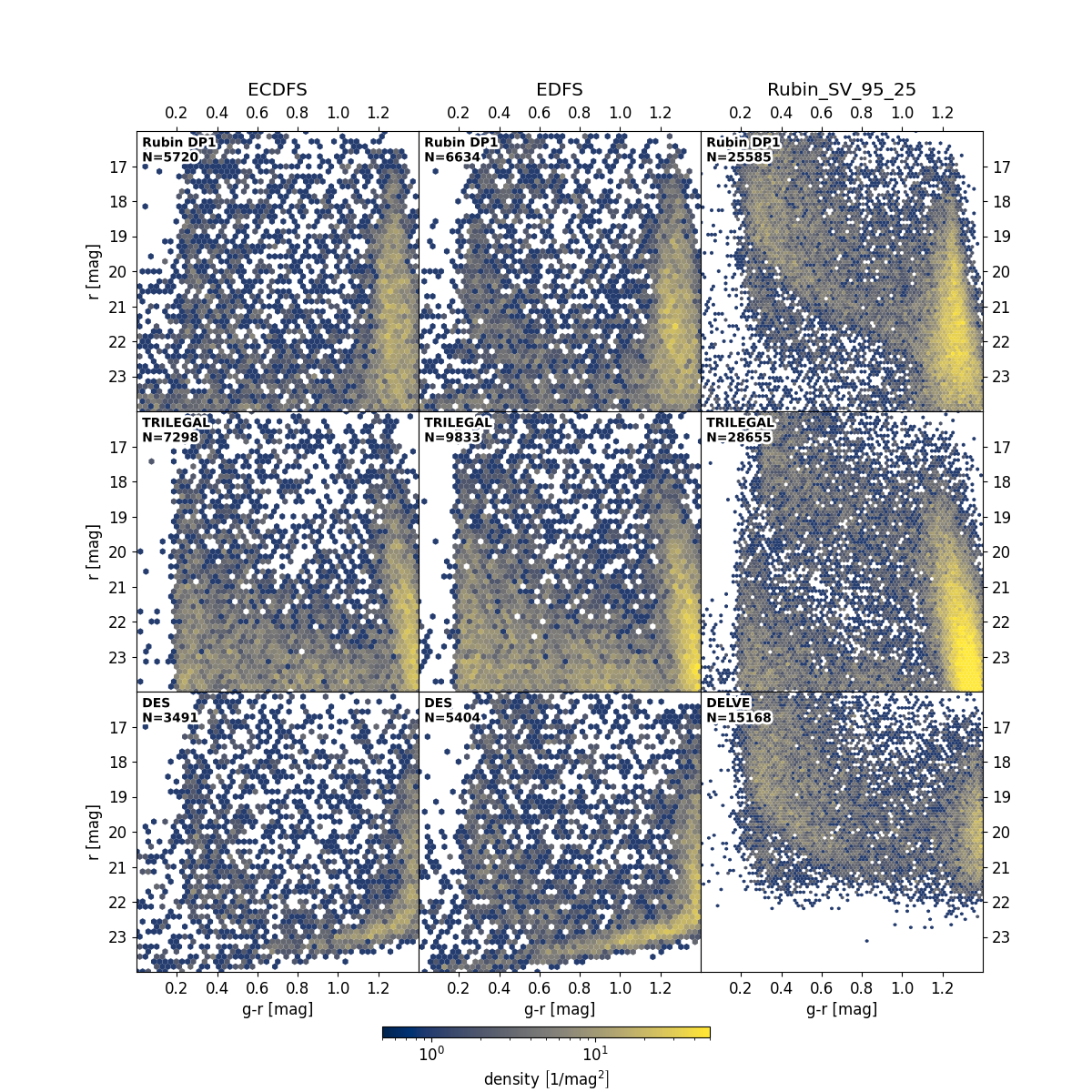}
	\caption{A comparison of color-magnitude diagrams of unresolved sources selected from Rubin DP1 (top row), TRILEGAL simulation (middle row) and DES and DELVE data (bottom row). Each column corresponds to one of the Rubin DP1 fields (as designated in the top left corner of each panel). Color scale is according to the probability density. While bin sizes are smaller in the rightmost column, the same color scale is shared between all panels. TRILEGAL, DELVE and DES data have been selected so that the area of each field is equal to the approximate area covered by the corresponding DP1 fields. Recall that $r<24$, star-galaxy separation and $0 < g-r < 1.4$ selection criteria were applied. DELVE data contain only a few stars fainter than \textit{r}$\approx$22. Please note a deficit of faint ($r>22$) blue stars ($g-r< 0.6$), compared to TRILEGAL simulation, in all three fields.}
	\label{fig:CMDs}
\end{figure*}

The top row in Figure~\ref{fig:CMDs} shows color-magnitude diagrams of unresolved sources selected from the three Rubin DP1 fields analyzed here. We select the plotted sample by limiting the apparent magnitude range to $r<24$ \footnote{Where there is no evidence for appreciable contamination by galaxies (recall Figure~\ref{fig:compareCMDs}).} and apply our star-galaxy separation method. Results of these cuts are listed in Table \ref{tab:fields}.  Additionally, we restrict the color range to $0 < g-r < 1.4$.

While stellar counts generally increase with magnitude $m$ (for a uniform source distribution proportionally to 10$^{0.6m}$,
with the fiducial case of flat counts corresponding to a volume number density decreasing as $1/d^3$ with the distance $d$) in
all three fields, there is a clear deficit of blue stars ($0.2 < g-r < 0.6$) at $r\sim22-24$, compared to stars with similar
colors at brighter magnitudes. 

\begin{figure}[h!]
	\includegraphics[width=0.495\textwidth,angle=0]{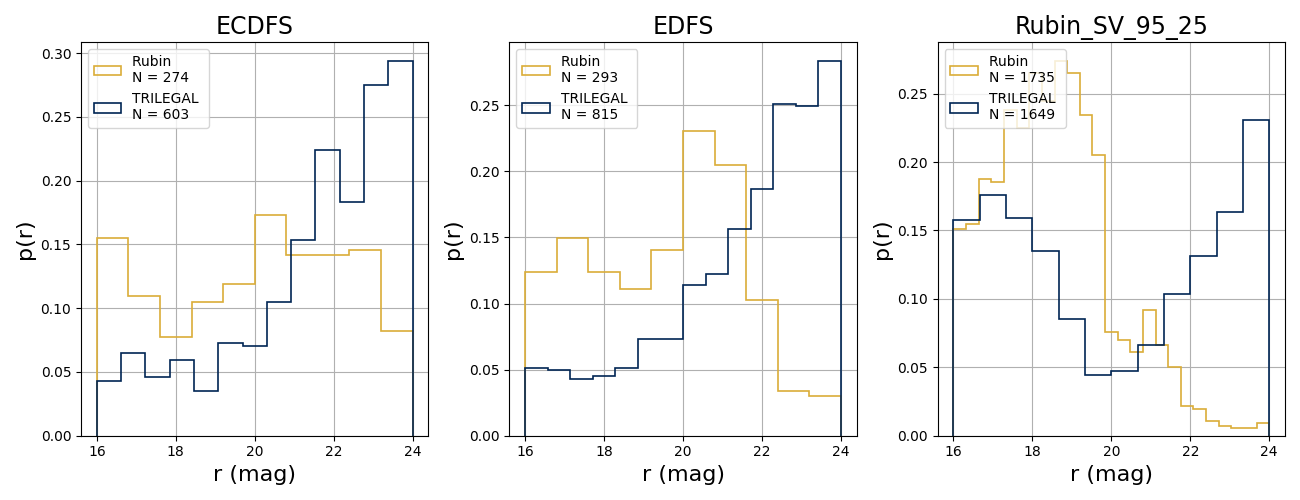} 
	\caption{A comparison of observed counts and TRILEGAL simulation using sources with $0.3 < g-r < 0.4$,
          separately for each of the three DP1 fields, as marked at the top of each panel. Note that TRILEGAL simulation 
          predicts steep rise of counts for $r< 21$, which is not seen in Rubin data.}
	\label{fig:rCountsTRILEGAL}
\end{figure}

\begin{figure}[h!]
	\includegraphics[width=0.495\textwidth,angle=0]{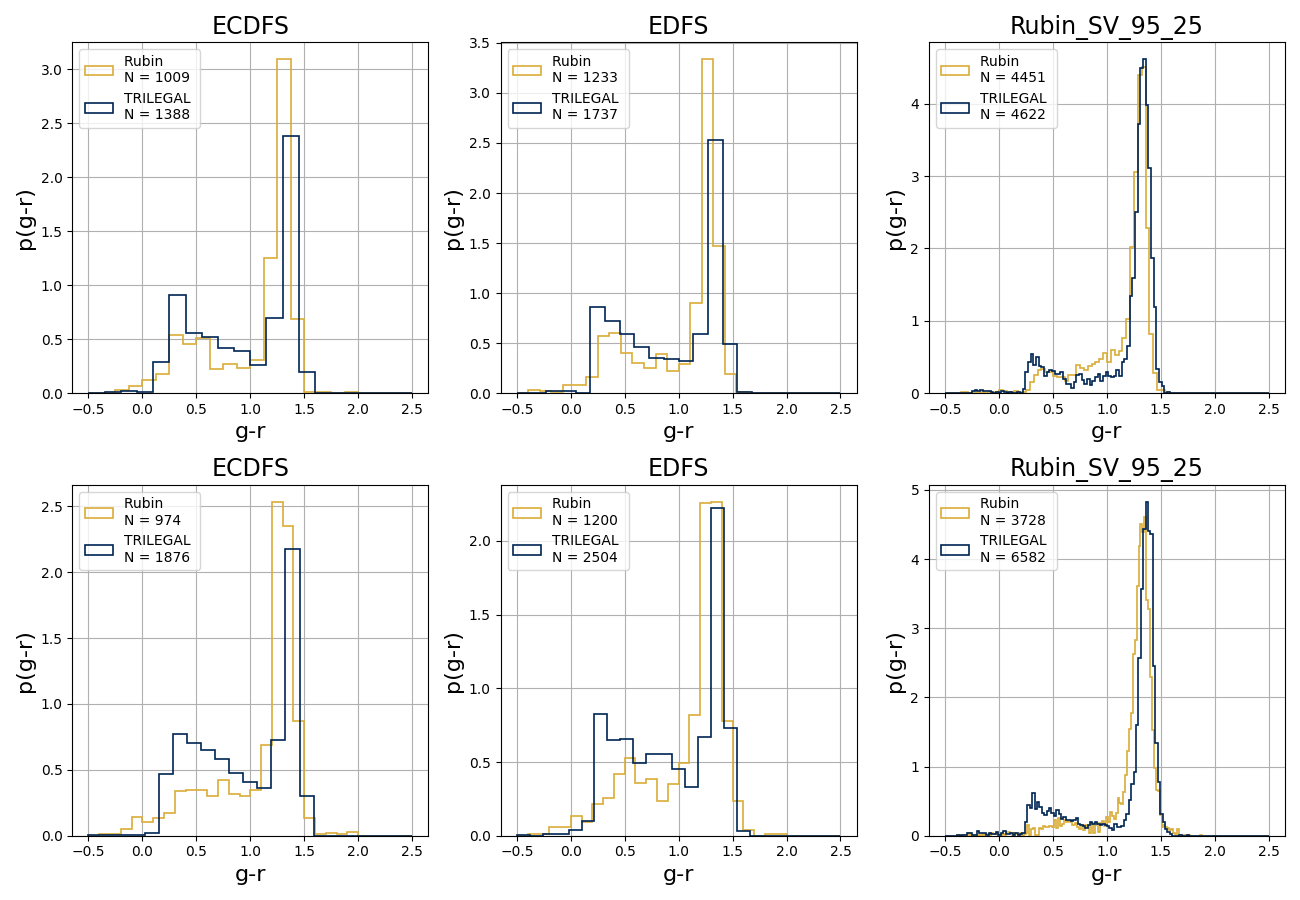}
	\caption{A comparison of observed $g-r$ color distributions and those predicted by TRILEGAL simulation
          for stars from two magnitude bins (top: $22<r<23$, bottom: $23<r<24$), 
          separately for each of the three DP1 fields, as marked at the top of each panel. 
          Note that TRILEGAL simulation predicts significantly more blue turn-off stars.}
	\label{fig:grhistTRILEGAL} 
\end{figure}

The middle row in Figure~\ref{fig:CMDs} shows color-magnitude diagrams predicted by TRILEGAL simulation.
It is evident that predicted counts of faint blue stars exceed the observed counts. It is remarkable that TRILEGAL simulation reproduces
the unusually well-defined separation of the disk and halo populations in the SV 95 $-$25 field: a diagonal paucity of stars
extending from ($r\sim19$, $g-r\sim0.2$) to ($r\sim22$, $g-r\sim0.6$). Figures~\ref{fig:rCountsTRILEGAL} and \ref{fig:grhistTRILEGAL} 
provide a more quantitative visualization of this discrepancy by comparing the $r$-band counts in a narrow color range
and the $g-r$ color histograms for two faint slices of the $r$ band magnitude. 

The region of the color-magnitude diagram where TRILEGAL overpredicts observed counts is dominated by halo stars.
Motivated by SDSS results \citep{2008ApJ...673..864J, 2010ApJ...714..663D}, TRILEGAL assumes the following halo number
density profile for the halo component 
\begin{equation}
        n(R, Z) = n_0 \left(R^2 + (Z/q)^2\right)^{-n/2},
\label{eq:rho}  
\end{equation}
where $R$ and $Z$ are galactocentric cylindrical coordinates, $n_0$ is a population-specific constant, $n=2.75$ is the halo power-law index and $q=0.62$ measures
the halo oblateness ($q=1$ for a spherical halo). This profile extends to a distance of 200 kpc, implying that counts for
blue main sequence turn-off stars should remain high to about $r\sim26.5$ (for more details, see \citealt{2020MNRAS.497.1547P}). 
The implication of the observed deficit of faint blue stars is that the power-law index is much larger and the resulting
profile much steeper at large galactocentric radii than observed by SDSS at brighter magnitudes. A more quantitative analysis
is presented in \S3.

DP1 is the very first release of Rubin photometric catalogs and all results need to be interpreted with caution since
it will take more time and more data to fully understand the behavior of image processing pipelines. For this reason,
we compare Rubin data in these three fields to extant datasets (see section~\ref{sec:extant}). The bottom row in Figure~\ref{fig:CMDs}
shows color-magnitude diagrams constructed using DES and DELVE catalogs. While these datasets appear a bit shallower
than DP1, it is evident that they support our conclusion regarding the significant deficit of faint blue stars. Additionally, tests based on artificial source injection show that the source detection incompleteness at faint magnitudes cannot explain the blue star deficit and small, barely resolved, galaxies, do not show any deficit but smooth increase of counts with magnitude \citep{RTN-095}. 

\subsection{Distribution of Stars in Rubin Color-color Diagrams and Comparison with SDSS}

For a more quantitative analysis of the stellar number density profile, estimates of stellar distances are required.
Stellar distances can be estimated from broad-band colors using the Bayesian framework discussed in
\cite{2025AJ....169..119P}. The stellar model color library required by the framework is derived from spectral energy
distribution models using the SDSS photometric system. We note that when computing likelihoods they express all colors as functions of $g-i$ color. 

Although SDSS and Rubin photometric systems are similar, we assess their differences by comparing the distribution of unresolved sources in Rubin's color-color diagrams to stellar model color sequences from \cite{2025AJ....169..119P} (which are in excellent agreement with SDSS
observations).

\begin{figure}[h!]
	\includegraphics[width=0.495\textwidth,angle=0]{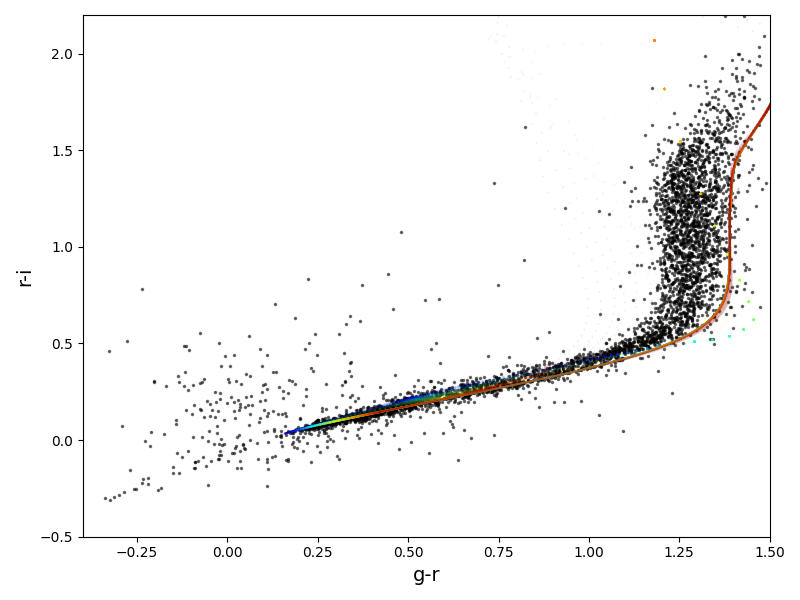}
	\caption{A comparison of the distribution of unresolved sources from Rubin DP1 in the $r-i$ vs. $g-r$ color-color diagram
          (symbols) and stellar model color sequences from \cite{2025AJ....169..119P}, which are in excellent agreement
          with SDSS observations. The sequences are color-coded by metallicity (-2.5 $<$ [Fe/H] $<$ +0.5, linearly from blue to red),  but in this color projection they are
          by and large degenerate (for more details see Figure 3 in \citealt{2025AJ....169..119P}). Note that M stars
          with $r-i > 0.5$ have a mean $g-r$ color of $\sim$1.3, while in the SDSS photometric system $g-r \sim$1.4.}
	\label{fig:grriDiag} 
\end{figure}

\begin{figure}[h!]
	\includegraphics[width=0.495\textwidth,angle=0]{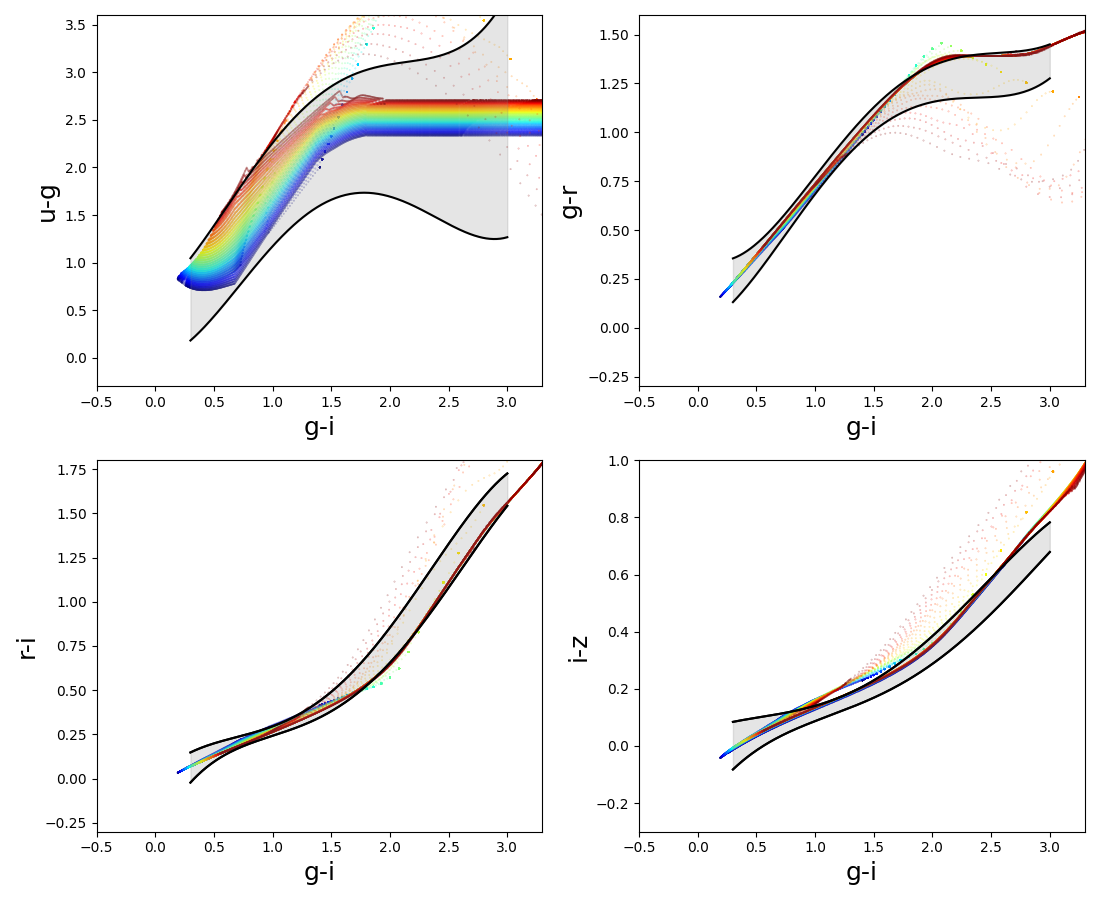}
	\caption{A comparison of the color distributions of unresolved sources in Rubin DP1 (the shaded gray regions
          bounded by solid black lines represent $\pm2\sigma$ envelope around the median color in $g-i$ bins)
          and stellar model color sequences from \cite{2025AJ....169..119P}. The sequences correspond to a
          10 Gyr population are color-coded by metallicity; the full metallicity range is $-2.5 < [Fe/H] < +0.5$ and coded linearly from blue to red (for additional details see Figure 3 in \cite{2025AJ....169..119P}).}
  	\label{fig:giColorComp}
\end{figure}

The largest discrepancy between SDSS and Rubin photometry for stars is seen in the $g-r$ color distribution for
red stars (M spectral type), as illustrated in Figure~\ref{fig:grriDiag}. Based on a preliminary analysis of DP1 stars
that also have SDSS photometry by M. Porter et al. (Rubin Technical Note RTN-099, in preparation), this difference
in the $g-r$ color comes primarily from the color term between the Rubin and SDSS $g$ bandpasses, due to Rubin's
$g$ bandpass red edge extending about 20 nm further to longer wavelengths compared to the SDSS $g$ bandpass. The second
largest color term is observed in the $z$ band, with a strength of about one third of that in the $g$ band.  The color term in $r$ band is much smaller. Additionally, we systematically compare Rubin data and the stellar models used by \texttt{photoD}, available in the SDSS photometric system in  Figure~\ref{fig:giColorComp}.

Unfortunately, there are no available constraints for color terms in the Rubin $u$ and $y$ bands. The $u$
band is especially important because it provides metallicity estimates, which in turn is needed for accurate
distance estimates. Based on the top left panel in Figure~\ref{fig:giColorComp}, it seems that there are no
large offsets (say, larger than 0.1 mag) in the $u-g$ color between the Rubin and SDSS systems. The
models in the $g-r$ vs. $u-g$ color-color diagram appear to be shifted upwards relative to Rubin data, but note
that at faint magnitudes probed by Rubin the majority of blue stars is expected to belong to the low-metallicity halo
population \citep{2008ApJ...684..287I}.  

\subsubsection{Selection of Main Sequence Stars using Stellar Locus Selection} 

Given the stellar locus parametrization in Rubin photometric system shown in Figure~\ref{fig:giColorComp},
the number of non-main-sequence stars (e.g., white dwarfs), quasars and unresolved galaxies in the unresolved
sample can be further decreased by rejecting objects with colors that are not consistent with the 5-dimensional
stellar color locus.  We reject all objects that are further than $2\sigma$ away from the median locus parametrized
by the $g-i$ color (that is, outside the gray regions shown in Figure~\ref{fig:giColorComp}. We note that red giant
stars can also be found in the main stellar locus -- however, their fraction is negligible at the faint magnitudes probed here. The rejected sources probably dominated by small unresolved galaxies and non-main-sequence stars with varying colors (e.g. unresolved binary stars).
We have repeated the analysis with a $4\sigma$ rejection cut but our results for the stellar number density profiles and
metallicity distribution did not appreciably change. 

\begin{figure}[h!]
	\includegraphics[width=0.495\textwidth,angle=0]{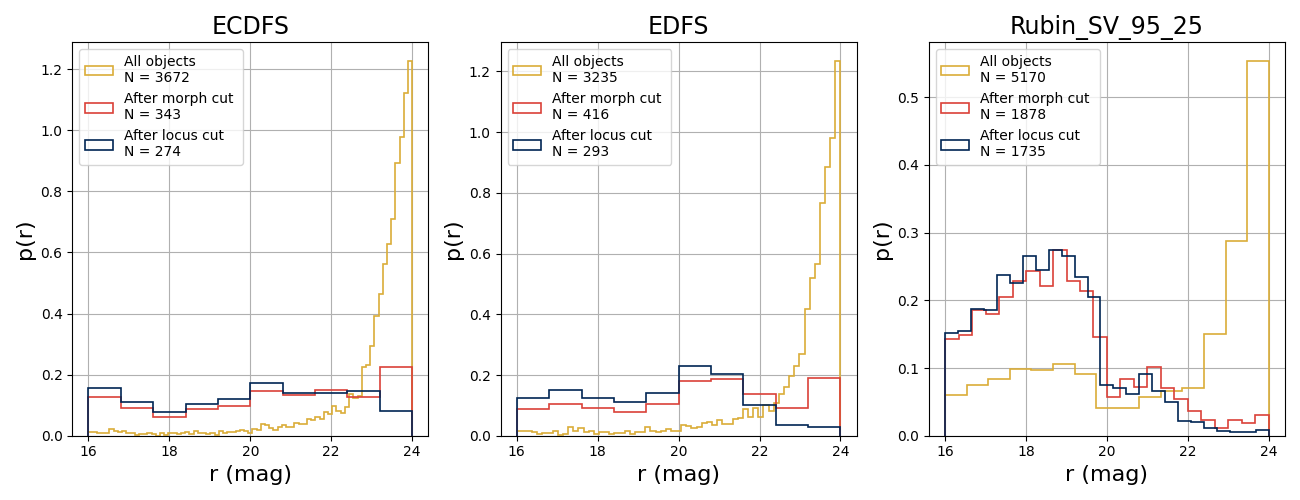} 
	\caption{A comparison of counts for selected subsamples using sources with $0.3 < g-r < 0.4$,
          separately for each of the three DP1 fields, as marked at the top of each panel. Note that in
          the $22 < r < 23$ magnitude range the effect of locus color cuts on the counts profile is negligible
          and remains minimal in the  $23 < r < 24$ magnitude range.}
	\label{fig:rCounts}
\end{figure}

Because we will analyze the stellar number density profile in the next section, it is important to ensure that this
additional color-based rejection does not significantly alter the completeness of faint blue stars in the sample.
As shown in Figure~\ref{fig:rCounts}, the effect of locus color cuts on the counts profile is negligible for
$r<23$ and remains minimal in the  $23 < r < 24$ magnitude range. The main effect and purpose of this
selection step is to reject sources for which distance estimates would not be reliable in any case
because their colors are inconsistent with the main-sequence stellar locus.

\subsection{Stellar Photometric Distance Estimation}

The Bayesian \texttt{photoD} framework compares observed colors to
a library of model colors for various stellar populations. For main sequence stars, the models are
parametrized by stellar luminosity (i.e., absolute magnitude, $M_r$), metallicity $[Fe/H]$ and
population age (of secondary importance when discussing data at high galactic latitudes as considered
here). The models are augmented by the dust extinction (reddening) along the line of sight, parametrized
by the extinction in the $r$ band ($A_r$). The framework thus provides the best-fit values of
$M_r$, $[Fe/H]$ and $A_r$ and their uncertainties.

Currently, there are two difficulties with applying this framework to the full Rubin DP1 dataset. First, the 
likelihood computation is very sensitive to discrepancies between observed and model colors. As we
discussed above, photometric transformations for all Rubin bands are not available yet. In addition,
priors derived from TRILEGAL simulations appear questionable at the faint end due to a deficit of observed
faint blue stars compared to model predictions. In order to obtain at least approximate estimates of
distances to halo stars, we instead follow \cite{2008ApJ...684..287I} and limit our analysis to blue stars near
the main sequence turn-off point ($g-i <1$). For these stars, $M_r$ can be determined using an analytic
expression for $M_r$ as a function of the $g-i$ color and $[Fe/H]$, with the metallicity estimated using the
$u-g$ and $g-r$ colors.

We estimate absolute magnitudes using eq.~A7 from \cite{2008ApJ...684..287I} and an updated expression
for metallicity from \cite{2010ApJ...716....1B} (eq. A1). Based on an analysis of stars in globular clusters,
\cite{2008ApJ...684..287I} estimated that the probable systematic errors in absolute magnitudes determined
using these relations are about 0.1 mag, corresponding to 5\% systematic distance errors (in addition
to the 10\%-15\% random distance errors). Since the DP1 dataset is at high galactic latitudes, we do not
fit for $A_r$ but instead adopt the extinction values from the dust maps by \cite{schlegel_maps_1998}. 

Based on the preliminary results \citep[Rubin Technical Note RTN-099, in. prep]{RTN-099}, we transform the Rubin's LSSTComCam $g-i$ color to SDSS-like $g-i$ color using \footnote{both valid in the range
$0.2 <  (g-i)_{ComCam} < 3.0$)}
\begin{equation}
     (g-i)_{SDSS} = 1.065 \, (g-i)_{ComCam} + 0.005,
   \end{equation}
and analogously for the $g-r$ color
\begin{equation}
     (g-r)_{SDSS} = 1.058 \, (g-r)_{ComCam} + 0.058 \, (r-i)_{ComCam} - 0.002. 
\end{equation}
In the $g-i$ range relevant here ($0.3 < g-i < 0.7$), the resulting $(g-i)_{SDSS}$ color is about 
0.03-0.05 mag redder than the $(g-i)_{ComCam}$ color, with a corresponding change of distance
scale by about 1.5-2.5\%. For the $g-r$ color, the above transformation properly reproduces
the mean SDSS $g-r \sim 1.4$ color for M stars (recall Figure~\ref{fig:grriDiag}). 

\begin{figure}[h!]
	\includegraphics[width=0.55\textwidth,angle=0]{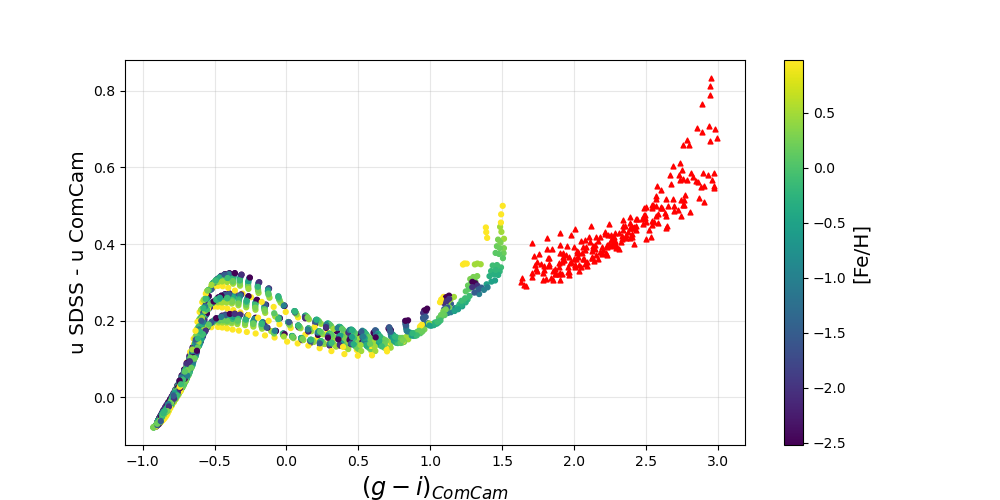}
	\caption{Predictions for the color term between Rubin and SDSS $u$ band magnitudes derived by integrating stellar model
	spectral energy distributions with photometric bandpasses. Symbols color-coded by metallicity (see the legend
	on the right) correspond to the Kurucz library models for main sequence stars with $log(g)=$ 4.0, 4.5 and 5.0 (effective
	temperature changes along the locus).  The red symbols correspond to models for cold stars from \cite{2013MSAIS..24..128A}.
	}
	\label{fig:uColorTerm}
\end{figure}

The $u$ band color term has the largest effect on distances via the impact of the $u-g$ color on $[Fe/H]$
estimates, and in turn their impact on $M_r$ estimates. For example, a systematic error in the $u-g$ color
of 0.02 mag induces a systematic $[Fe/H]$ errors ranging from 0.02 dex for high-metallicity stars to 0.11 dex
for low-metallicity halo stars \citep{2008ApJ...684..287I}. Unfortunately, the $u$ band color term is
currently not well known. Predictions based on stellar spectral energy distribution models (see Figure~\ref{fig:uColorTerm})
predict 0.15 mag around $g-i = 0.5$ relevant for blue stars considered below (due to the fact that the effective
wavelength for the Rubin $u$ bandpass is longer than for the SDSS $u$ band). There is no overlap between DP1
fields and SDSS sky coverage. A preliminary performance analysis based on unpublished Rubin commissioning
data of the COSMOS field with LSST Camera by the Rubin commissioning team supports the above model-based
predictions for blue stars but we note that the LSST Camera $u$ band filter and LSSTComCam $u$ band filter
are physically different devices. Motivated by the model predictions and empirical comparison with the COSMOS field,
we adopt the following simple corrections for blue stars with $g-i \sim 0.5$ considered below:
\begin{equation}
     u_{SDSS} = u_{ComCam} + 0.15.
     \label{eq:uColorTerm}  
\end{equation}
Its main effects are to shift metallicity distribution to higher values by about 0.4 dex, and to make $M_r$ brighter
by 0.2-0.3 mag, resulting in larger distances by 10-15\%. We return to this point and the effects of no $u$ band
correction in \S4. 

\section{Analysis of the Stellar Halo Number Density and Metallicity Profiles \label{sec:analysis}}

Here we construct and analyze the resulting halo number density profiles and metallicity distributions. We only consider stars
bluer than $g-r=0.60$ to ensure reliable $[Fe/H]$ estimates \citep{2008ApJ...684..287I} and further limit them to a
narrow absolute magnitude range, $4.0 < M_r < 5.5$ (approximately  corresponding to the $0.3 < g-i < 0.6$ color range),
to have a good control of the selection effects at the faint end. Assuming a 100\% completeness to the faint limit at $r=24$,
this subsample is complete to a distance modulus of 18.5 (50 kpc), with the completeness dropping to zero at
a distance modulus of 20 (100 kpc). The distribution of a selected subsample in galactocentric cylindrical coordinates is shown
in Figure~\ref{fig:RZ}. 

\subsection{Halo number density profiles}

\begin{figure}[h!]
\includegraphics[width=0.495\textwidth,angle=0]{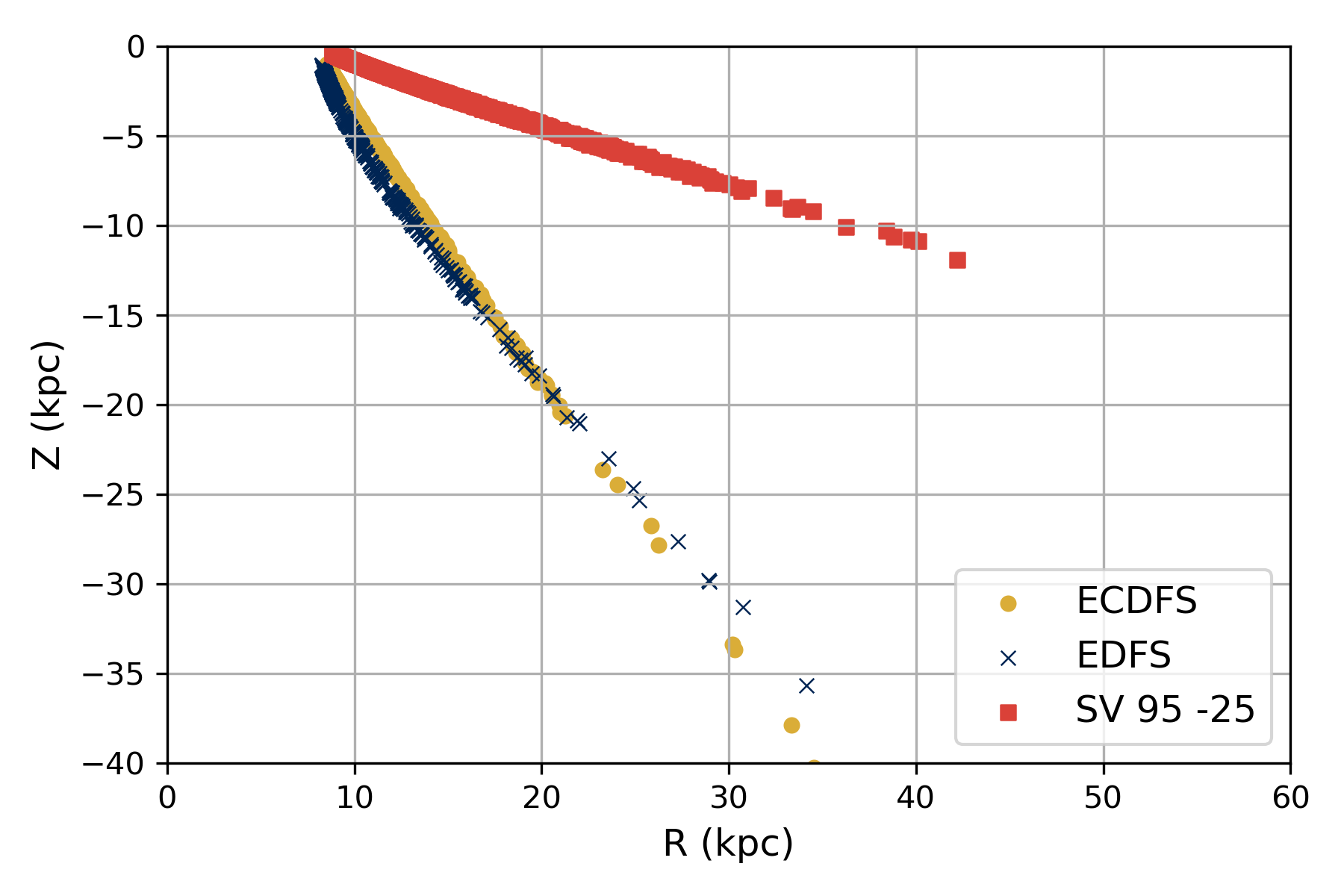}
\caption{The distribution of selected blue stars in galactocentric cylindrical coordinates for the three DP1 fields as
  marked in the legend. The starting sample sizes are listed in the last column in Table \ref{tab:fields}. After further
  cuts ($g-r<0.6$ and $4.0 < M_r < 5.5$), the three subsamples include 378 (ECDFS), 387 (EDFS) and 3452 (SV 95 -25) stars. 
  Note the poor coverage of the $R-Z$ plane with only three pencil beams.}
\label{fig:RZ}
\end{figure}

\begin{figure}[h!]
\includegraphics[width=0.495\textwidth,angle=0]{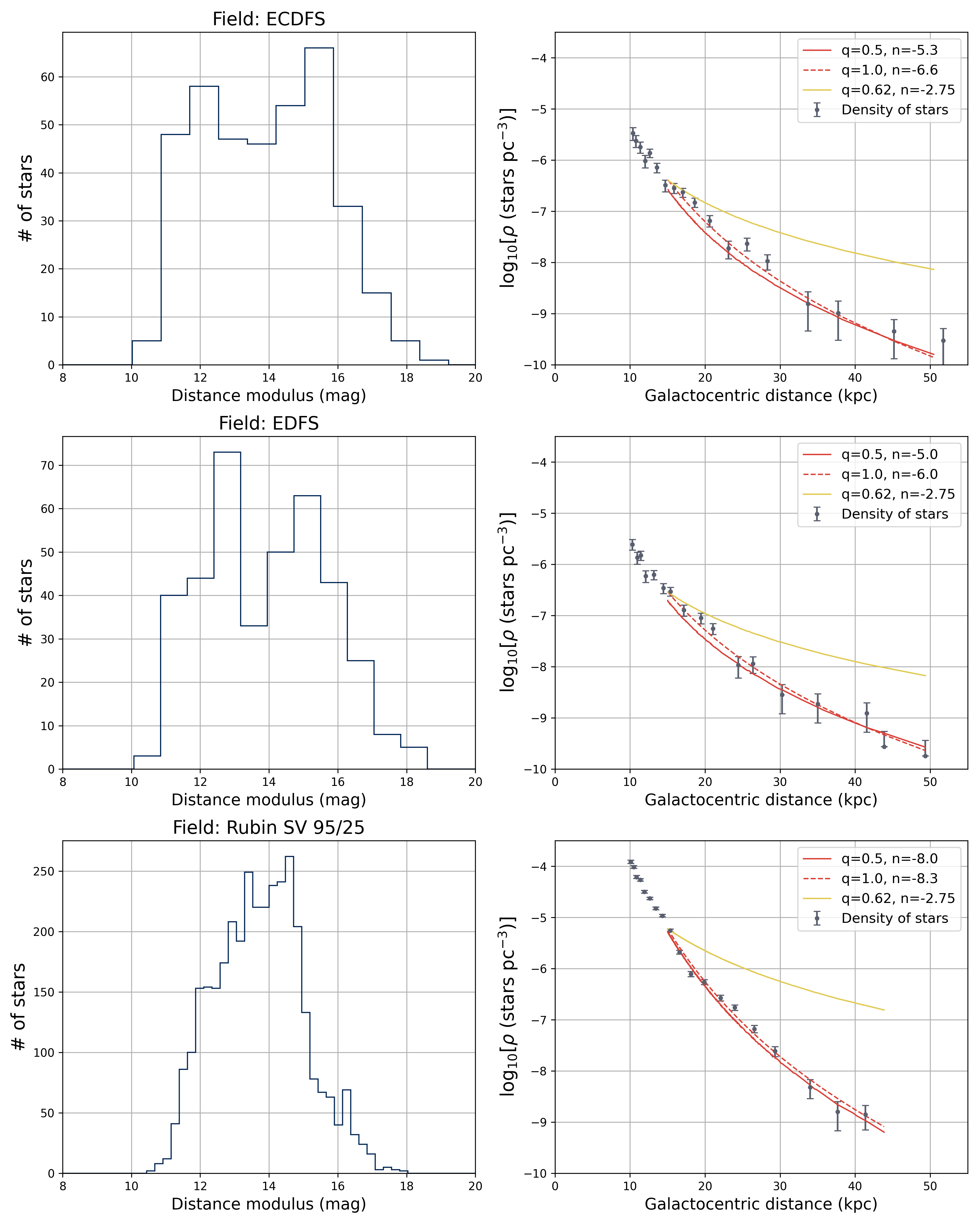}
\caption{Each row shows the distance modulus distribution (left) and stellar number density profile (right) for the three Rubin DP1
  fields analyzed in this study. Note that the sample distance modulus completeness limit is 18.5 and thus the steeply decreasing
  right edge of the distance modulus distributions is a real effect and not a selection effect. In the right column, data are shown as
  symbols with Poisson uncertainties and lines are axially symmetric elliptical halo models (Eq.~\ref{eq:rho}), with the corresponding
  $n$ and $q$ parameters listed in the legend (see text). Note that the data display much steeper profiles than the SDSS-motivated
  $q=0.62$ and $n=2.75$ halo profile assumed in TRILEGAL simulations.}
\label{fig:density_result}
\end{figure}

Using stellar positions shown in Figure~\ref{fig:RZ}, we compute their number density profile. For each field separately,
we bin the distance modulus (see the left panels in Figure~\ref{fig:density_result}) and for each bin compute the corresponding
number density by dividing the stellar counts in a bin by the bin volume (easily computed as the volume difference of two cones
defined by distance modulus bin boundaries). The resulting number density profiles are shown in the right panels in
Figure~\ref{fig:density_result}.  

We fit axially symmetric elliptical halo models (Eq.~\ref{eq:rho}) to data at galactocentric distances beyond 15 kpc, where
halo counts dominate over the counts of disk stars. The best-fit profiles, with $n$ ranging from 5 to 8, are much
steeper than the SDSS-motivated $q=0.62$ and $n=2.75$ halo profile, measured at  distances below 10 kpc and assumed in
TRILEGAL simulations. Given the poor coverage of the $R-Z$ plane with only three pencil beams, it is not possible to break
the degeneracy between $q$ and $n$ and we provide best-fit $n$ values for assumed values of $q=0.5$ and $q=1$. 
Therefore, the best-fit values $n$ should not be overinterpreted. 

\begin{figure}[h!]
\includegraphics[width=0.495\textwidth,angle=0]{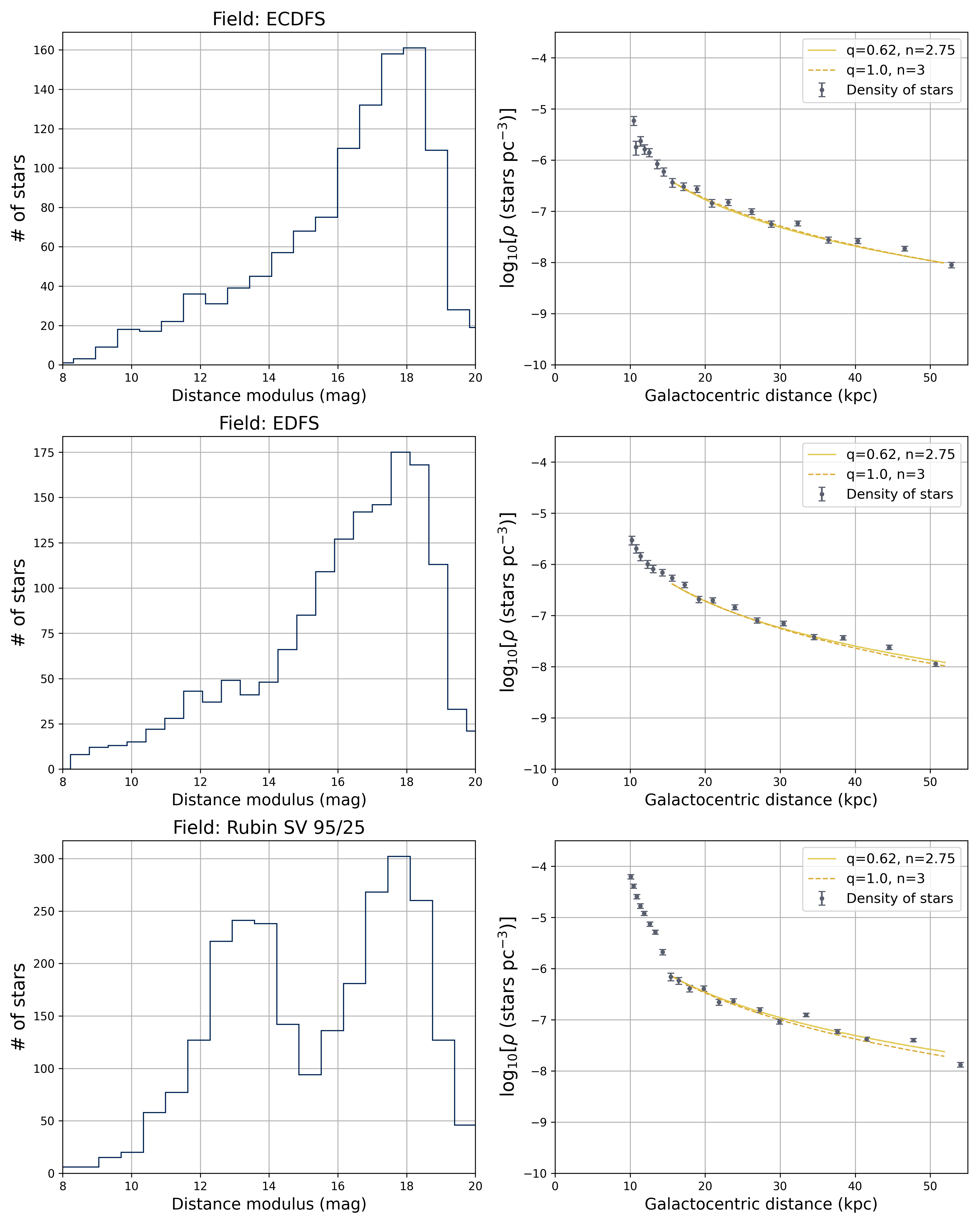}
\caption{Analogous to Figure~\ref{fig:density_result}, except that the TRILEGAL simulation is used instead of Rubin DP1 data.
 Note that the $q=0.62$ and $n=2.75$ halo profile assumed in simulations is faithfully reproduced (though it cannot be
 distinguished from the canonical spherical $1/r^3$ profile due to sparse coverage of the $R-Z$ plane).}
\label{fig:trilegal_comp}
\end{figure}

As a test of our analysis code, we repeat identical steps using simulated TRILEGAL catalog. As evident in Figure~\ref{fig:trilegal_comp},
the $q=0.62$ and $n=2.75$ halo profile assumed in TRILEGAL simulations is faithfully reproduced for all three sky directions.
It is interesting to compare the bottom left panels in Figures~\ref{fig:density_result} and \ref{fig:trilegal_comp}. The
paucity of predicted halo stars for distance modulus values larger than 17 is clearly seen (and directly related to the results shown in
Figures~\ref{fig:rCountsTRILEGAL} and \ref{fig:grhistTRILEGAL}).

\subsection{Halo metallicity distributions}

\begin{figure}[h!]
\includegraphics[width=0.495\textwidth,angle=0]{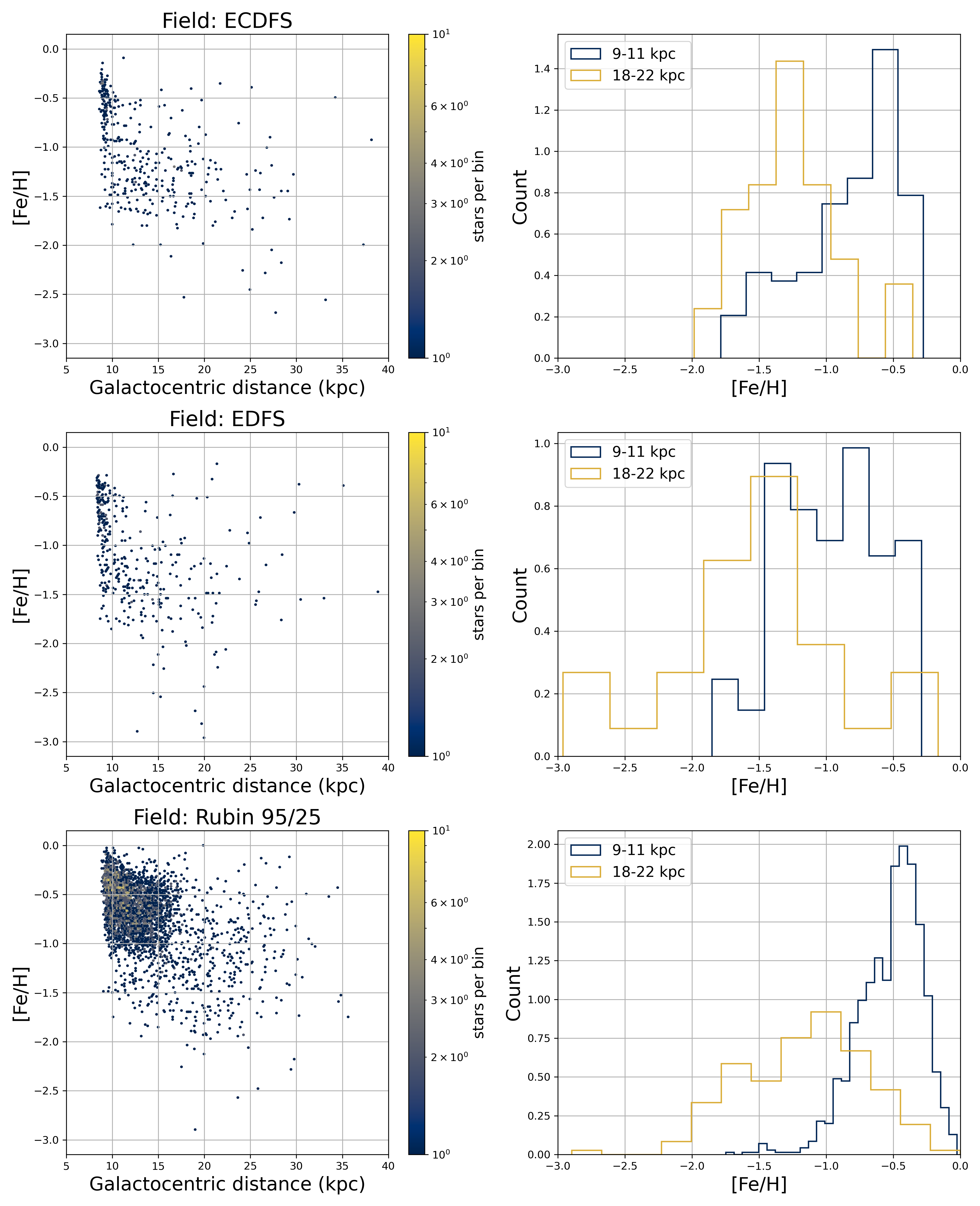}
\caption{Each row shows the variation of stellar metallicity with galactocentric distance (left) and metallicity distribution in two
  slices of distance (right), as marked in the legend. See text for discussion.}
\label{fig:feh_vs_d}
\end{figure}

\begin{figure}[h!]
\includegraphics[width=0.495\textwidth,angle=0]{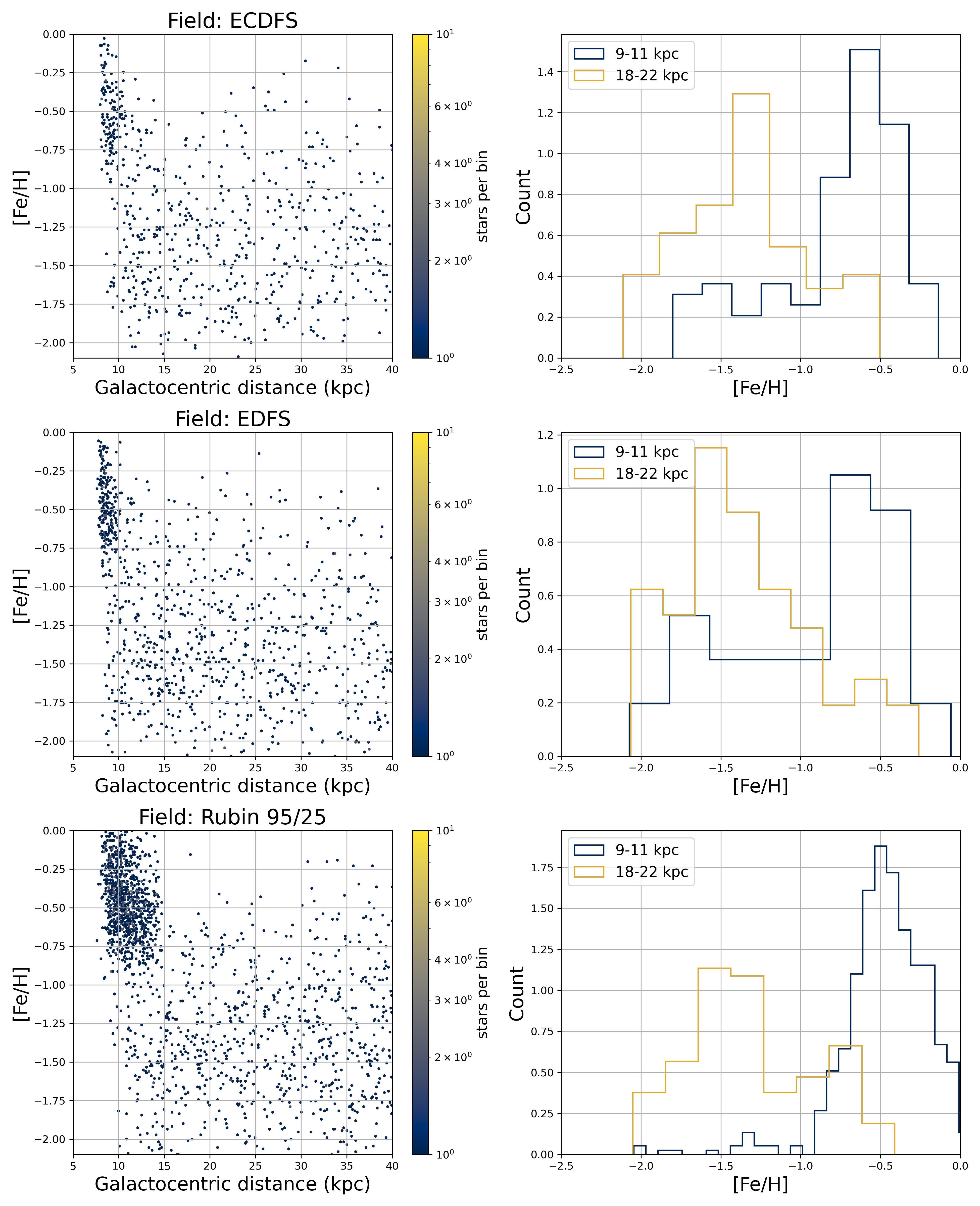}
\caption{Analogous to Figure~\ref{fig:feh_vs_d}, except that the TRILEGAL simulation is used instead of Rubin DP1 data.
 Note that the metallicty distribution in the more distant bin dominated by halo stars is consistent with a distribution centered 
 on $[Fe/H]= -1.5$ assumed in simulations.}
\label{fig:feh_vs_d_trilegal}
\end{figure}

The metallicity distribution of selected blue stars is shown in Figure~\ref{fig:feh_vs_d}. The distributions in the $\sim$20
kpc galactocentric distance slice are consistent with expectations for halo stars (the halo metallicity distribution for stars 
within 10 kpc measured by SDSS is centered on $[Fe/H] = -1.5$; \citealt{2008ApJ...684..287I}). The 10 kpc bin is more
representative of high-metallicity disk stars. For comparison, and again as a test of our analysis code, Figure~\ref{fig:feh_vs_d_trilegal} 
shows results obtained with the simulated TRILEGAL catalog. As expected, the metalicity distribution in the 20 kpc bin is consistent with the
assumed distribution centered on $[Fe/H] = -1.5$, while the 10 kpc bin seems dominated by higher-metallicity disk stars,
similar to observed DP1 distributions. 

\section{Discussion and Conclusions \label{sec:disc}}

Distances to stars are a crucial ingredient in our quest to better understand the formation and evolution
of the Milky Way galaxy. Anticipating photometric catalogs with tens of billions of stars from Rubin's LSST,
we investigated the utility of Rubin's DP1 catalogs for estimating stellar number density profile in the Milky Way halo.
Although DP1 includes only three very small fields, it nevertheless enabled several important results presented here. 

We found a strong deficit of faint ($r>22$) blue main sequence stars compared to cutting-edge TRILEGAL simulations,
a finding that is also supported by extant data from the DES and DELVE surveys. This discrepancy is interpreted as a signature of
a much steeper halo number density profile at galactocentric distances $10-50$ kpc than the SDSS-motivated $1/r^{2.75}$
profile assumed in TRILEGAL simulations. This interpretation is consistent with earlier suggestions based on observations
of more luminous, but much less numerous, evolved stellar populations, and a few pencil beam surveys of blue main
sequence stars in the northern sky: at about 20-30 kpc, the profile slope changes from about $n\sim2-3$ to about $n\sim4-5$
(see Figure 7 in \citealt{2024MNRAS.531.4762M} for a visual summary of results from many studies).  We were unable to break
the degeneracy between the oblateness parameter ($q$) and the exponent $n$ given only three pencil beam constraints from DP1
dataset. For this reason, the measured exponents $n$ should be interpreted with caution (for a related discussion, see \citealt{2015A&A...579A..38P}).

Our metallicity results agree with expectations of a halo distribution centered on $[Fe/H] = -1.5$, as measured by SDSS out to 10 kpc.
Nevertheless, the applied $u$ band color term (eq.~\ref{eq:uColorTerm}) of 0.15 mag is probably uncertain to about 0.05 mag.
As a result of that uncertainty, there are likely systematic errors in observed $[Fe/H]$ distributions (shifts along the abscissa) of about 0.2
dex (and implied distance scale changes of about 10\%). We note that uncertainty in the $u$ band correction has no discernible
effect on the halo profile shape since it is well described by a power law. Therefore, we cannot exclude a halo metallicity
gradient of up to about 0.2 dex between galactocentric distances of 10 kpc and 50 kpc.

As Rubin collects more data with the much larger LSST Camera than LSSTComCam in sky regions with SDSS coverage, it will be possible to derive
robust and precise photometric transformations between the Rubin and SDSS photometric systems (e.g., the equatorial
SDSS Stripe 82 region would be perfect for such studies, for details see \citealt{2021MNRAS.505.5941T}). In addition,
the LSST photometric throughput will be well measured and can be used to compute colors from
stellar spectral energy distribution models without an intermediate reference to the SDSS system. With improved stellar model colors, and with
adequate adjustments to stellar population models, such as TRILEGAL, used for priors, it will be
possible to deploy the full Bayesian framework described in \cite{2025AJ....169..119P}. 

Ultimately, Rubin's LSST will cover half the sky, an area more than 10,000 times larger than, for example, the ECDFS field
analyzed here, and to a similar depth. As one among many benefits of a large sky coverage, LSST data will break the
$q-n$ degeneracy when fitting halo number density profiles. Further anticipated improvements in this context brought
about by the 10-year LSST dataset are discussed in detail by \cite{2025AJ....169..119P}. We conclude that the results
presented here bode well for future explorations of the Milky Way with LSST.

\section*{Acknowledgements}
        
   This work was supported by the Croatian Science Foundation under the project number IP-2025-02-1942 and has been financed within the Tenure Track Pilot Programme of the Croatian Science Foundation and the Ecole Polytechnique Fédérale de Lausanne and the Project TTP-2018-07-1171 "Mining the Variable Sky", with the funds of the Croatian-Swiss Research Programme.
   
   This material is based upon work supported in part by the National
   Science Foundation through Cooperative Agreements AST-1258333
   and AST-2241526 and Cooperative Support Agreements AST-1202910 and
   2211468 managed by the Association of Universities for
   Research in Astronomy (AURA), and the Department of Energy under
   Contract No. DE-AC02-76SF00515 with the SLAC National
   Accelerator Laboratory managed by Stanford University. Additional
   Rubin Observatory funding comes from private donations,
   grants to universities, and in-kind support from LSST-DA Institutional Members.

   L.P. acknowledges support from LSST-DA through grant 2024-SFF-LFI-08-Palaversa. This work was conducted as part of a LINCC Frameworks Incubator. LINCC Frameworks
   is supported by Schmidt Sciences. W.B., D.B, A. J. C., N.C., S.C., M.D., D.J., O.L., K.M., A.I.M., and S.M., are supported by Schmidt Sciences.
      
   \v{Z}.I, M.J. and N.C. acknowledge support from the DiRAC Institute in the Department of Astronomy at the University of Washington. M.J. and N.C. would also like to acknowledge support by the National Science Foundation under Grant No. AST-2003196. 
   
   The DiRAC Institute is supported through generous gifts from the Charles and
   Lisa Simonyi Fund for Arts and Sciences and the Washington Research
   Foundation. Funding for the SDSS and SDSS-II has been provided by the Alfred
   P. Sloan Foundation, the Participating Institutions, the National
   Science Foundation, the U.S. Department of Energy, the National
   Aeronautics and Space Administration, the Japanese Monbukagakusho, the
   Max Planck Society, and the Higher Education Funding Council for
   England. The SDSS Web Site is \url{https://www.sdss.org/}.
   
   GP, LG, MT and SZ acknowledge support from project "SPICE4LSST - Adding incompleteness and crowding errors to the LSST Data Releases" (INAF Large Grant 2024).
   LG and SZ acknowledge support from "Population synthesis with rotating stars: a necessary upgrade" (INAF Theory Grant 2022).
   MT and GP acknowledge financial support by the European Union – NextGenerationEU and by the University of Padua under the 2023 STARS Grants@Unipd programme (``CONVERGENCE: CONstraining the Variability of Evolved Red Giants for ENhancing the Comprehension of Exoplanets'').
    
   This work has made use of data from the European Space Agency (ESA)
   mission Gaia \url{https://www.cosmos.esa.int/gaia}, processed by the Gaia
   Data Processing and Analysis Consortium (DPAC,
   \url{https://www.cosmos.esa.int/web/gaia/ dpac/consortium}). Funding for the
   DPAC has been provided by national institutions, in particular, the
   institutions participating in the Gaia Multilateral Agreement.
   
   \section*{Facilities}
   Vera C. Rubin Observatory, Gaia, 
   Sloan Digital Sky Survey, Blanco Telescope and DECam
   
   \section*{Software}
   \texttt{Astropy} \citep{astropy-1, astropy-2},
   \texttt{AstroML} \citep{2012cidu.conf...47V},
   \texttt{HATS} \citep{2025arXiv250102103C},
   \texttt{Jupyter} \citep{jupyter},
   \texttt{LSDB} \citep{2025arXiv250102103C},
   \texttt{Matplotlib} \citep{matplotlib},
   \texttt{Numpy} \citep{numpy},
   \texttt{Scipy} \citep{scipy},
   \texttt{Seaborn} \citep{seaborn},
   \texttt{Pandas} \cite{pandas},
   \texttt{Python} \citep{python}

   \bibliographystyle{aa} 
   \bibliography{paper} 

\end{document}